# Kepler-62: A five-planet system with planets of 1.4 and 1.6 Earth-radii in the Habitable Zone

William J. Borucki[*,1], Eric Agol[12], Francois Fressin[10], Lisa Kaltenegger[10,39], Jason Rowe[6], Howard Isaacson[2], Debra Fischer[19], Natalie Batalha[1], Jack J. Lissauer[1], Geoffrey W. Marcy[2], Daniel Fabrycky[17, 42], Jean-Michel Désert[10], Stephen T. Bryson[1], Thomas Barclay[35], Fabienne Bastien[43], Alan Boss[4], Erik Brugamyer[8], Lars A. Buchhave[40, 41], Chris Burke[6], Douglas A. Caldwell[6], Josh Carter[18], David Charbonneau[10], Justin R. Crepp[20,38], Jørgen Christensen-Dalsgaard[7], Jessie L. Christiansen[6], David Ciardi[22], William D. Cochran[8], Edna DeVore[6], Laurance Doyle[6], Andrea K. Dupree[10], Michael Endl[8], Mark E. Everett[14], Eric B. Ford[16], Jonathan Fortney[17], Thomas N. Gautier III[11], John C. Geary[10], Alan Gould[13], Michael Haas[1], Christopher Henze[1], Andrew W. Howard[24], Steve B. Howell[1], Daniel Huber[32], Jon M. Jenkins[6], Hans Kjeldsen[7], Rea Kolbl[2], Jeffery Kolodziejczak[25], David W. Latham[10], Brian L. Lee[12], Eric Lopez[17] , Fergal Mullally[6], Jerome A. Orosz[30], Andrej Prsa[26], Elisa V. Quintana[6], Dimitar Sasselov[10], Shawn Seader[6], Avi Shporer[20,36], Jason H. Steffen[9], Martin Still[35], Peter Tenenbaum[6], Susan E. Thompson[6], Guillermo Torres[10], Joseph D. Twicken[6], William F. Welsh[30], Joshua N. Winn[18]

[*]To whom correspondence should be addressed. Email: William.J.Borucki@nasa.gov

[1]NASA Ames Research Center, Moffett Field, CA 94035, USA.
[2]University of California, Berkeley, CA, 94720, USA.
[3]San Jose State University, San Jose, CA, 95192, USA.
[4]Carnegie Institution of Washington, Washington, DC 20015 USA.
[6]SETI Institute, Mountain View, CA, 94043, USA.
[7]Department of Physics & Astronomy, Aarhus University, Aarhus, Denmark.
[8]McDonald Observatory, University of Texas at Austin, Austin, TX, 78712, USA.
[9]Northwestern University, Evanston, IL, 60208, USA.
[10]Harvard-Smithsonian Center for Astrophysics, Cambridge, MA, 02138, USA.
[11]Jet Propulsion Laboratory, Calif. Institute of Technology, Pasadena, CA, 91109, USA.
[12]Department of Astronomy, Box 351580, University of Washington, Seattle, WA 98195, USA.
[13] Lawrence Hall of Science, Berkeley, CA 94720, USA.
[14]NOAO, Tucson, AZ 85719 USA.
[15]University of Arizona, Steward Observatory, Tucson, AZ 85721, USA.
[16]Univ. of Florida, Gainesville, FL, 32611 USA.
[17]Univ. of Calif., Department of Astronomy and Astrophysics, Santa Cruz, CA 95064 USA.
[18]MIT, Cambridge, MA 02139 USA.
[19]Yale University, New Haven, CT 06520 USA.
[20]California Institute of Technology, Department of Astronomy, Pasadena, CA 91125 USA.
[21]Orbital Sciences Corp., Mountain View, CA 94043 USA.
[22]Exoplanet Science Institute/Caltech, Pasadena, CA 91125 USA.
[23]University Affiliated Research Center, University of California, Santa Cruz, CA 95064 USA.
[24]Institute for Astronomy, University of Hawaii, Honolulu, HI 96822, USA.
[25]MSFC, Huntsville, AL 35805 USA.
[26]Villanova University, Villanova, PA 19085 USA.
[30]San Diego State University, San Diego, CA 92182, USA.

[34]University of Amsterdam, Amsterdam, The Netherlands.
[35]Bay Area Environmental Research Institute/ Moffett Field, CA 94035, USA.
[36]Las Cumbres Observatory Global Telescope, Goleta, CA 93117, USA.
[37]Solar System Exploration Division, NASA Goddard Space Flight Center, Greenbelt, MD 20771, USA.
[38]University of Notre Dame, Dept. of Physics, Notre Dame, IN 46556, USA.
[39]Max Planck Institute of Astronomy, Koenigstuhl 17, 69115 Heidelberg, Germany.
[40]Niels Bohr Institute, University of Copenhagen, DK-2100, Copenhagen, Denmark.
[41]Centre for Star and Planet Formation, Natural History Museum of Denmark, University of Copenhagen, DK-1350 Copenhagen, Denmark.
[42]Department of Astronomy and Astrophysics, University of Chicago, Chicago, IL 60637, USA.
[43]Vanderbilt University, Nashville, TN 37235, USA.

## Abstract

**We present the detection of five planets -- Kepler-62b, c, d, e, and f -- of size 1.31, 0.54, 1.95, 1.61 and 1.41 Earth radii ($R_\oplus$), orbiting a K2V star at periods of 5.7, 12.4, 18.2, 122.4 and 267.3 days, respectively. The outermost planets (Kepler-62e & -62f) are super-Earth-size (1.25 < planet radius ≤ 2.0 $R_\oplus$) planets in the habitable zone (HZ) of their host star, receiving 1.2 ± 0.2 and 0.41 ± 0.05 times the solar flux at Earth's orbit ($S_\odot$). Theoretical models of Kepler-62e and -62f for a stellar age of ~7 Gyr suggest that both planets could be solid: either with a rocky composition or composed of mostly solid water in their bulk.**

## Main Text

Kepler is a NASA Discovery-class mission designed to determine the frequency of Earth-radius planets in and near the HZ of solar-like stars (*1-6*). Planets are detected as "transits" that cause the host star to appear periodically fainter when the planets pass in front it along the observer's line of sight. Kepler-62 (KIC 9002278, KOI 701) is one of approximately 170,000 stars observed by the Kepler spacecraft. Based on an analysis of long-cadence photometric observations from Kepler taken in Quarters 1 through 12 (May 13, 2009 through March 28, 2012), we report the detection of five planets including two super-Earth-size planets in the HZ and a hot Mars-size planet orbiting Kepler-62 (Fig. 1 and Table 1). Prior to validation, three of these objects were designated as planetary candidates KOI-701.01, 701.02, and 701.03 in the Kepler 2011 catalog (*7*) and the Kepler 2012 catalog (*8*). KOI-701.04 and 701.05 were identified subsequently using a larger data sample (*9*).

Analysis of high-resolution spectra indicates that Kepler-62 is a K2V spectral type with an estimated mass and radius (in solar units) of 0.69 ± 0.02 $M_\odot$ and 0.63 ± 0.02 $R_\odot$ (*9*). Examination of the sky close to Kepler-62 showed the presence of only one additional star that contributed as much as 1% to the total flux (figs. S3-S4)(*9*). Warm-Spitzer observations (fig. S9) and the analysis of centroid motion (Table S1) were consistent with the target star as the source of the transit signals (Fig. 1 and fig. S1). We computed the radius, semi-major axis, and radiative equilibrium temperature of each planet (Table 1) based on light curve modeling given the derived stellar parameters (Table S3).

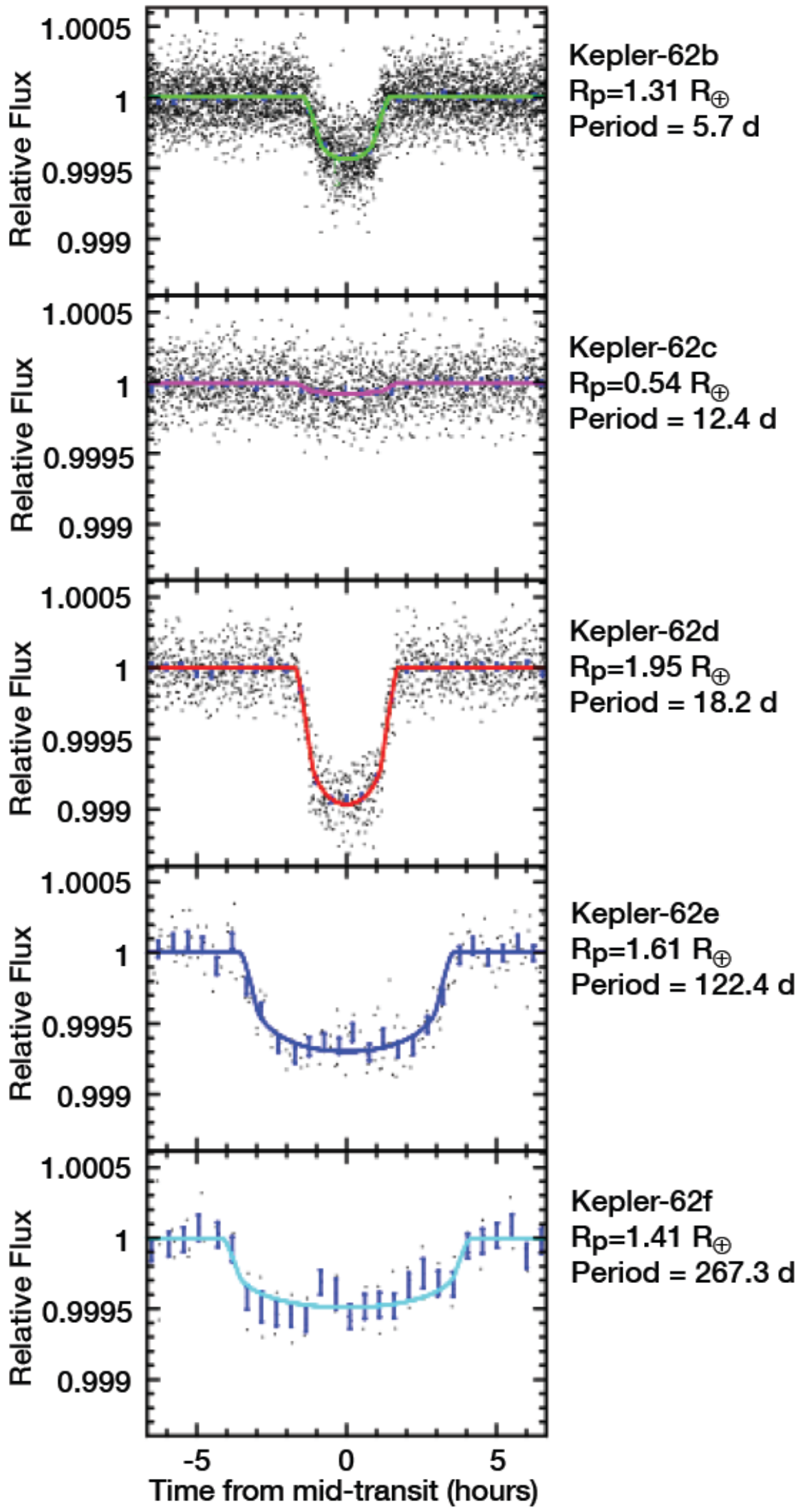

Fig. 1. Kepler-62 light curves after the data were detrended to remove the stellar variability. Composite of phase-folded transit light curves (dots), data binned in ½ hour intervals (blue error bars), and model fits (colored curves) for Kepler-62b through -62f. Model parameters are provided in Table 1. The error bars get larger as the period becomes larger because there are fewer points to bin together. For the shortest periods, the bars are too small to see.

**Table 1.** Characteristics of the Kepler-62 planetary system.

| Parameter | Kepler-62b | Kepler-62c | Kepler-62d | Kepler-62e | Kepler-62f |
|---|---|---|---|---|---|
| $T_0$ (BJD-2454900) | 103.9189 ± 0.0009 | 67.651 ± 0.008 | 113.8117 ± 0.0008 | 83.404 ± 0.003 | 522.710 ± 0.006 |
| $P$ [days] | 5.714932 ± 0.000009 | 12.4417 ± 0.0001 | 18.16406 ± 0.00002 | 122.3874 ± 0.0008 | 267.291 ± 0.005 |
| duration [hr] | 2.31 ± 0.09 | 3.02 ± 0.09 | 2.97 ± 0.09 | 6.92 ± 0.16 | 7.46 ± 0.20 |
| depth [%] | 0.043 ± 0.001 | 0.007 ± 0.001 | 0.092 ± 0.002 | 0.070 ± 0.003 | 0.042 ± 0.004 |
| $R_p/R_*$ | 0.0188 ± 0.0003 | 0.0077 ± 0.0004 | 0.0278 ± 0.0006 | 0.0232 ± 0.0003 | 0.0203 ± 0.0008 |
| $a/R_*$ | 18.7 ± 0.5 | 31.4 ± 0.8 | 40.4 ± 1.0 | 144 ± 4 | 243 ± 6 |
| $b$ | 0.25 ± 0.13 | 0.16 ± 0.09 | 0.22 ± 0.13 | 0.06 ± 0.05 | 0.41 ± 0.14 |
| $i$ | 89.2 ± 0.4 | 89.7 ± 0.2 | 89.7 ± 0.3 | 89.98 ± 0.02 | 89.90 ± 0.03 |
| $e\cos\omega$ | 0.01 ± 0.17 | -0.05 ± 0.14 | -0.03 ± 0.24 | 0.05 ± 0.17 | -0.05 ± 0.14 |
| $e\sin\omega$ | -0.07 ± 0.06 | -0.18 ± 0.11 | 0.09 ± 0.09 | -0.12 ± 0.02 | -0.08 ± 0.10 |
| $a$ [AU] | 0.0553 ± 0.0005 | 0.0929 ± 0.0009 | 0.120 ± 0.001 | 0.427 ± 0.004 | 0.718 ± 0.007 |
| $R_p$ [$R_⊕$] | 1.31 ± 0.04 | 0.54 ± 0.03 | 1.95 ± 0.07 | 1.61 ± 0.05 | 1.41 ± 0.07 |
| Maximum mass ($M_⊕$)(*9*) | 9 | 4 | 14 | 36 | 35 |
| *Number of observed transits* | 171 | 76 | 52 | 8 | 3 |
| *Total SNR* | 54 | 8.5 | 68 | 31 | 12 |
| Radiative equilibrium temperature (K) | 750 ± 41 | 578 ± 31 | 510 ± 28 | 270 ± 15 | 208 ± 11 |

Notes: 1) $T_0$ is the epoch in mid-transit in Barycentric Julian Days, $P$ is the period, duration is the transit duration, "depth" is the percent reduction of the flux during the transits determined from the model fit to the data, $R_p/R_*$ is the ratio of the radius of the planet to the radius of the star, $a/R_*$ is the ratio of the planet's semi-major axis to the stellar radius, $b$ is the impact parameter in units of stellar radius, $i$ is the orbital inclination, $e\cos\omega$ is the product of the orbital eccentricity $e$ with the cosine of the periapse angle $\omega$, $a$ is the semi-major axis, and $R_p$ is the radius of the planet, and Maximum Mass is the upper limit to the mass based on transiting timing and RV observations, $M_⊕$ is the mass of the Earth, and Teq is the radiative equilibrium temperature.

2) The values of the uncertainties are ±1 standard deviation unless otherwise noted.

3) Values for the maximum mass are for the 95$^{th}$ percentile. See (*9*).

4) A second set of values for the planetary parameters was computed by an independent model and found to be in good agreement with the listed values.

The masses of the planets could not be directly determined using radial velocity (RV) measurements of the host star because of the planets' low masses, the faintness and variability of the star, and the level of instrument noise. In the absence of a detected signal in the RV measurements (*9*, Section 5), we statistically validate the planetary nature of Kepler-62b through -62f with the BLENDER procedure (*10-13*) by comparing the probability of eclipsing binaries and other false positive scenarios to bona-fide transiting planet signals (*14-18*).

We performed a systematic exploration of the different types of false positives that can mimic the signals, by generating large numbers of synthetic light curves that blend together light from multiple stars/planets over a wide range of parameters and comparing each blend with the Kepler photometry (Fig. 2). We rejected blends that result in light curves inconsistent with the observations. We then estimated the frequency of the allowed blends by taking into account all available observational constraints from the follow-up observations discussed in (*9*). Finally we compared this frequency with the expected frequency of true planets (planet "prior") to derive the "odds ratio" (*9*, §7).

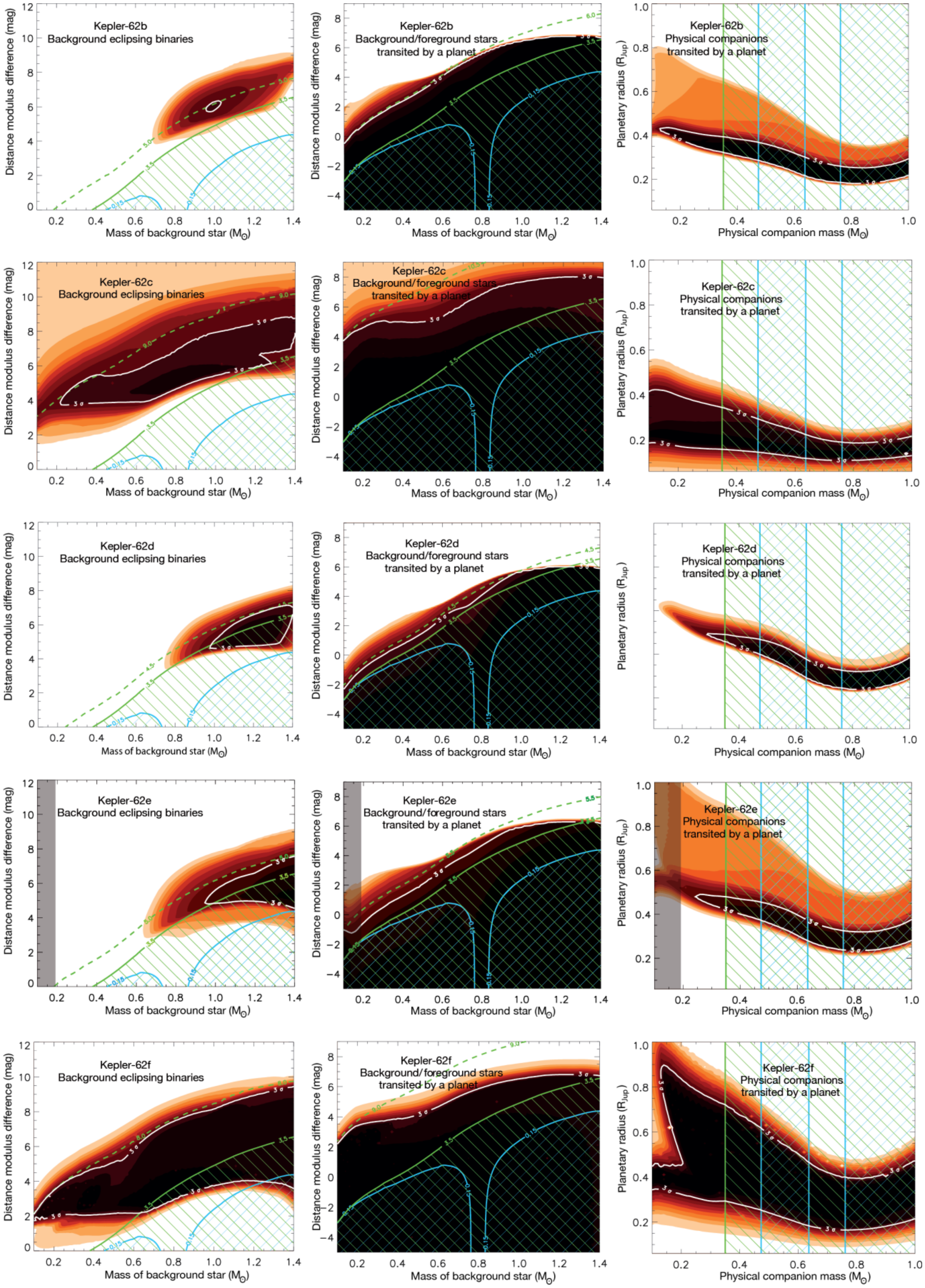
Kepler-62b
Background eclipsing binaries
Kepler-62b
Background/foreground stars
transited by a planet
Kepler-62b
Physical companions
transited by a planet
Kepler-62c
Background eclipsing binaries
Kepler-62c
Background/foreground stars
transited by a planet
Kepler-62c
Physical companions
transited by a planet
Kepler-62d
Background eclipsing binaries
Kepler-62d
Background/foreground stars
transited by a planet
Kepler-62d
Physical companions
transited by a planet
Kepler-62e
Background eclipsing binaries
Kepler-62e
Background/foreground stars
transited by a planet
Kepler-62e
Physical companions
transited by a planet
Kepler-62f
Background eclipsing binaries
Kepler-62f
Background/foreground stars
transited by a planet
Kepler-62f
Physical companions
transited by a planet
Distance modulus difference (mag)
Planetary radius (R_Jup)
Mass of background star (M_⊙)
Physical companion mass (M_⊙)

**Fig. 2.** BLENDER goodness-of-fit contours for Kepler-62bcdef corresponding to the three different scenarios that contribute to the overall blend frequency: background eclipsing binaries (left column), background or foreground stars transited by a planet (middle column), and physical companions transited by a planet (right column). Viable blends must be less than about 5.0 (left) or 5.5 (middle) magnitudes fainter than Kepler-62 (dashed line). Only blends inside the solid white contour match the Kepler light curve within acceptable limits (3σ, where σ is the significance level of the chi-square difference compared to a transit model fit). Lighter-colored areas (red, orange, yellow) mark regions of parameter space giving increasingly worse fits to the data (4σ, 5σ, etc.), and correspond to blends that we consider to be ruled out. The cyan cross-hatched areas indicate regions of parameter space that we consider ruled out because the resulting r-Ks color of the blend is either too red (left) or too blue (right) compared to the measured color, by more than 3σ (0.15 mag). The green hatched regions indicate blends that are ruled out because the intruding stars are less than 3.5 magnitudes fainter than the target and would be so bright that they would have been detected spectroscopically. Finally, the thin gray area on the left panel for Kepler-62e rule out stars based on our Spitzer observations (fig. S8), (*9*, §2.3). The likelihood of a false positive for each planetary candidate is derived from the integration of the area that remains within the 3σ boundary that is not eliminated by the hatched areas.

Incorporating these constraints into a Monte Carlo (MC) model that considers a wide range of stellar and planetary characteristics provides estimates of the probabilities of a false positive that could explain the observations (*9*, §6).

Our simulations of each of the candidates indicate that the likelihood of a false-positive explanation is much smaller than the likelihood of the planetary system explanation. In particular, the calculated odds ratios that Kepler-62b through -62f represent planets rather than false-positives are 5400, >5000, 15000, 14700, and > 5000, respectively (*9*, §7). There is also a 0.2% chance that the planets orbit a widely space binary composed of two K2V stars and therefore the planets are √2 larger in radius than shown in Table 1 (*9*, §7).

To determine if a planet is in the HZ, we calculated the flux of stellar radiation that it intercepts. It is convenient to express intercepted flux in units of the average solar flux intercepted by Earth, denoted by $S_{\odot}$. The values of the stellar flux intercepted by Kepler-62b to -62f are 70 ± 9, 25 ± 3, 15 ± 2, 1.2 ± 0.2 and 0.41 ± 0.05 $S_{\odot}$. Eccentric planetary orbits increase the annually averaged irradiation from the primary star by a factor of $1/(1-e^2)^{1/2}$ (*19*). Because the model results for the orbital eccentricities of Kepler-62b through -62f are small and consistent with zero, no corrections were made.

The HZ is defined here as the annulus around a star where a rocky planet with a $CO_2/H_2O/N_2$ atmosphere and sufficiently large water content (such as on Earth) can host liquid water on its solid surface (*20*). In this model, the locations of the two edges of the HZ are determined based on the stellar flux intercepted by the planet and the assumed composition of the atmosphere. A conservative estimate of the range of the HZ (labeled "narrow HZ" in Fig. 3) is derived from atmospheric models by assuming that the planets have a $H_2O$– and $CO_2$–dominated atmosphere with no cloud feedback (*21*). The flux range is defined at the inner edge by thermal run-away due to saturation of the atmosphere by water vapor and at the outer edge by the freeze-out of $CO_2$. In this model the planets are assumed to be geologically active and that climatic stability is provided by a mechanism in which atmospheric $CO_2$ concentration varies inversely with planetary surface temperature.

The "empirical" HZ boundaries are defined by the solar flux received at the orbits of Venus and Mars at the epochs when they potentially had liquid water on their surfaces. Venus and Mars are believed to have lost their water at least 1 Gyr and 3.8 Gyr ago, respectively when the Sun was less luminous. At these epochs, Venus received a flux of 1.78 $S_\odot$ and Mars a flux of 0.32 $S_\odot$ (*20*). The stellar-spectral-energy distributions of stars cooler than the Sun are expected to slightly increase the absorbed flux (*20*). Including this factor changes the HZ flux limits to 1.66 and 0.27 $S_\odot$ for the empirical HZ and 0.95 and 0.29 $S_\odot$ for the narrow HZ (*21*). Figure 3 shows that the Earth and Kepler-62f are within the flux-boundaries of the "narrow" HZ while Kepler-22b and Kepler-62e are within the "empirical" flux-boundaries.

Although RV observations were not precise enough to measure masses for Kepler-62e and -62f, other exoplanets with a measured radius below 1.6 $R_\oplus$ have been found to have densities indicative of a rocky composition. In particular, Kepler-10b (22), Kepler-36b (23), CoRoT-7b (24) have radii of 1.42 $R_\oplus$, 1.49 $R_\oplus$, 1.58 $R_\oplus$ and densities of 8.8, 7.5, and 10.4 gr/cc, respectively. Thus it is possible that both Kepler-62e and -62f (with radii of 1.61 $R_\oplus$ and 1.41 $R_\oplus$) are also rocky planets.

The albedo and the atmospheric characteristics of these planets are unknown, and therefore the range of equilibrium temperatures $T_{eq}$ at which the thermal radiation from each planet balances the insolation is large and depends strongly on the composition and circulation of the planets' atmospheres, their cloud characteristics and coverage, as well as the planets' rotation rates (*25, 26*). However, for completeness, values of $T_{eq}$ were computed from $T_{eq} = T_{\mathrm{eff}}\,[\beta\ (1\text{-}A_B)(R_*/2a)^2]^{1/4}$, where $T_{\mathrm{eff}}$ is the effective temperature of the star (4925°K), $R_*$ is the radius of the star (0.64), $A_B$ is the planet Bond albedo, $a$ is the planet semi-major axis, $\beta$ is a proxy for day-night redistribution with 1 for full redistribution and 2 for no redistribution. For the MCMC calculations, it was assumed that $\beta = 1$, and that $A_B$ is a random number from 0 to 0.5. (Table 1)

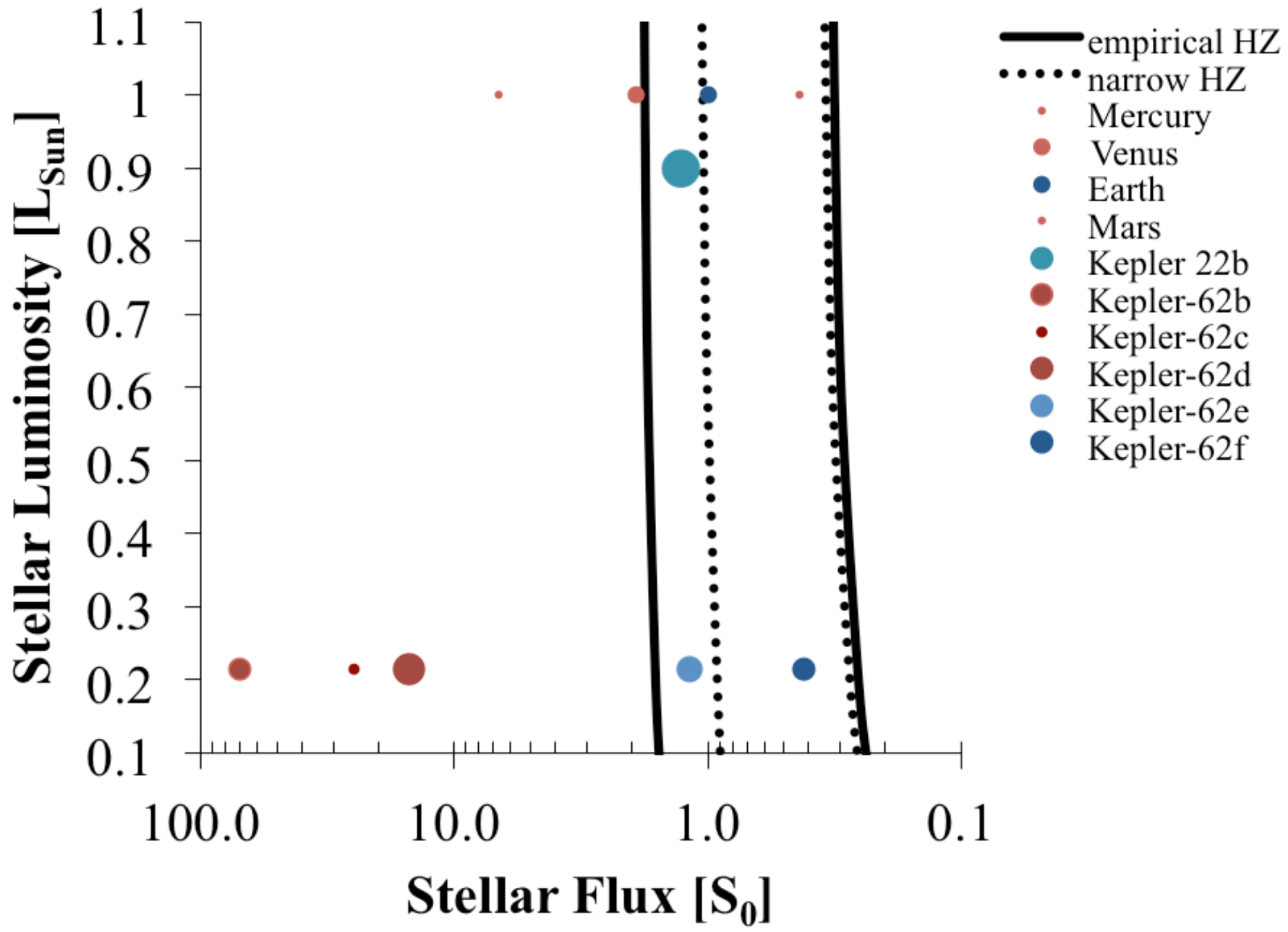

**Fig. 3.** Comparison of known exoplanets with measured radii less than 2.5 $R_{\oplus}$ in the HZ to the Solar System planets. The sizes of the circles indicate the relative sizes of the planets to each other. The dashed and the solid lines indicate the edges of the narrow and empirical HZ, respectively.

Gravitational interactions between Kepler-62e and -62f are too weak (*9*, §4) to cause non-linear variations in the times of transits (*27, 28*) and thereby provide estimates of their masses. Nevertheless, upper limits (95$^{th}$ percentile) for -62e and -62f were derived (Table S4): 150 $M_{\oplus}$ and 35 $M_{\oplus}$, respectively. A lower upper limit to the mass of Kepler-62e based on RV observations (Table S4) gives 36 $M_{\oplus}$. These values confirm their planetary nature without constraining their composition. Despite the lack of a measured mass for Kepler-62e and -62f, the precise knowledge of their radii, combined with estimates of their $T_{eq}$ and the stellar age (~7 Gyr) imply that Kepler-62e and -62f have lost their primordial or outgassed hydrogen envelope (*29, 30*). Therefore Kepler-62e and -62f are Kepler's first HZ planets that could plausibly be composed of condensable compounds and be solid, either as a dry, rocky super-Earth or one composed of a significant amount of water (most of which would be in a solid phase due to the high internal pressure) surrounding a silicate-iron core.

We do not know if Kepler-62e and -62f have a rocky composition, an atmosphere, or water. Until we get suitable spectra of their atmospheres we cannot determine whether they are in fact habitable. With radii of 1.61 and 1.41 $R_{\oplus}$ respectively, Kepler-62e and -62f are the smallest transiting planets detected by the Kepler Mission that orbit within the HZ of any star other than the Sun.

## Acknowledgements

Kepler was competitively selected as the tenth Discovery mission. Funding for this mission is provided by NASA's Science Mission Directorate. Some of the data presented herein were obtained at the W. M. Keck Observatory, which is operated as a scientific partnership among the California Institute of Technology, the University of California, and the National Aeronautics and Space Administration. The Keck Observatory was made possible by the generous financial support of the W. M. Keck Foundation. Lisa Kaltenegger acknowledges support from DFG funding ENP Ka 3142/1-1 and NAI. Funding for the Stellar Astrophysics Centre is provided by The Danish National Research Foundation. The research is supported by the ASTERISK project (ASTERoseismic Investigations with SONG and Kepler) funded by the European Research Council (Grant 267864). W. F. Welsh and J. A. Orosz acknowledge support from NASA through the Kepler Participating Scientist Program and from the NSF via grant AST-1109928. D. Fischer acknowledges support from NASA ADAP12-0172. O. R. Sanchis-Ojeda & J. N. Winn are supported by the Kepler Participating Scientist Program (PSP) through grant NNX12AC76G. E. Ford is partially supported by NASA PSP grants NNX08AR04G & NNX12AF73G. Eric Agol acknowledges NSF Career grant AST-0645416We would also like to thank the Spitzer staff at IPAC and in particular; Nancy Silbermann for checking and scheduling the Spitzer observations. The Spitzer Space Telescope is operated by the Jet Propulsion Laboratory, California Institute of Technology under a contract with NASA. The authors would like to thank the many people who gave so generously of their time to make this Mission a success.
All data products are available to the public at the Mikulski Archive for Space Telescopes; http://stdatu.stsci.edu/kepler/

**Supplementary Materials**
www.sciencemag.org/cgi/content/full/science.
Materials and Methods
SOM text, SOM Figs. S1-S14, SOM Tables S1-S4

# Supporting Online Material

**In this supplementary material, we provide additional details regarding the detection of five planets orbiting Kepler-62 (KIC 9002278, KOI-701). This supplement is organized as follows; In §1, we describe the data, the transit detection, and the image analysis used to detect signals from nearby stars. In §2 we describe the follow-up observations that help to further rule out non-planetary astrophysical sources. §3 describes the procedures to measure the host star characteristics including, effective temperature, surface gravity, metallicity, mass, age, and rotation rate. §4 discusses the search for transit timing variations, §5 describes planetary mass constraints based on high-precision Doppler measurements. §6 presents the MCMC light-curve modeling that yields estimates of the orbital and planet characteristics. §7 provides details of the validation of the planet interpretation.**

## 1. Candidate Identification

### 1.1 Data

The analysis presented here utilizes twelve quarters of 30-minute cadence data (Q1-Q12) spanning 1013.86 days between 13 May 2009 and 17 March 2012 (fig. S1). The duty cycle for cadences averages 93% of the elapsed time. However, the removal of the cadences taken near the thermal transients associated with monthly data downloads and during the transits of other Kepler-62 planets, causes the duty cycle to drop to 81%. The data were processed with various versions of the data analysis pipeline: Q1-Q8 with Pipeline version SOC 8.1, Q9-Q11 with Pipeline version SOC 8.0, and Q12 with Pipeline version SOC 8.1. For a description of each pipeline version, see the Kepler Data Handbooks at the Mikulski Archive at Space Telescope Institute (http://archive.stsci.edu/Eepler/Kepler_fov/search.php).

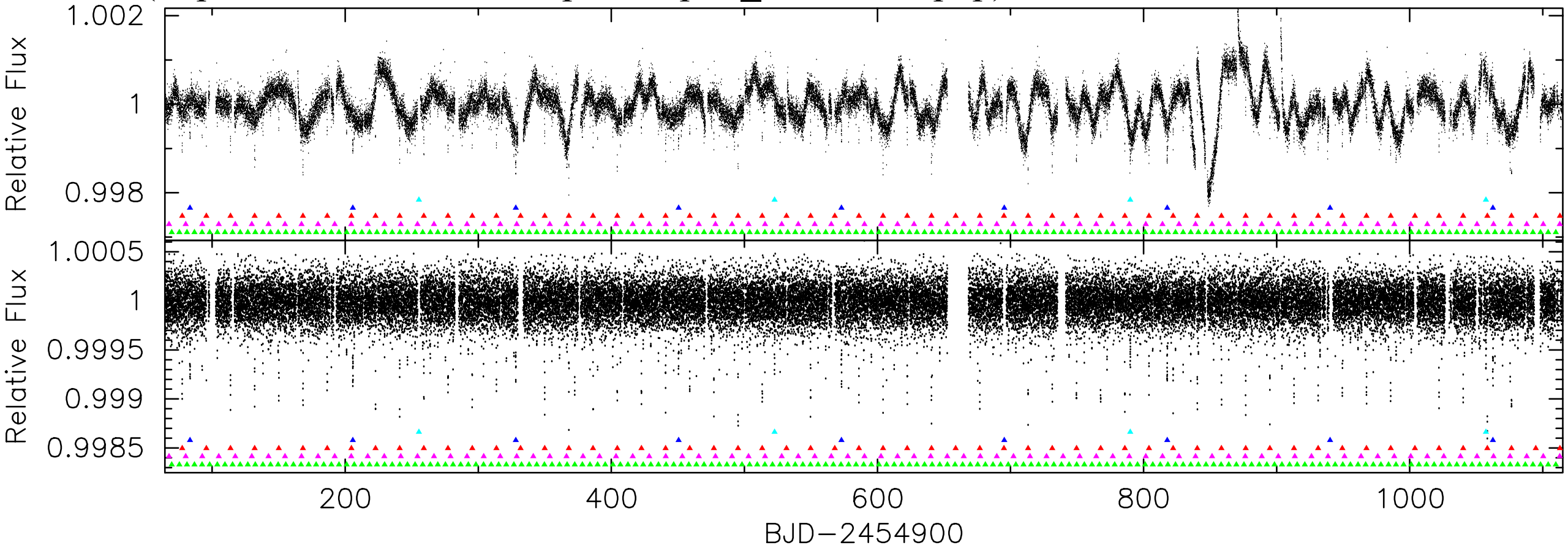

**Fig. S1.** Top panel; Q1-Q12 photometric time series after correction for trends and instrument systematics. It is a representation of the stellar variability. Bottom panel; detrended and normalized time series flux after the removal of both the instrument artifacts and stellar variability. It was used to model the planetary transits. The green, magenta, red, blue and cyan lines mark the occurrence of transits from Kepler-62b through -62f, respectively.

1.2 Transit detection

Kepler-62b, -62d, & -62e correspond to KOI-701.02, 701.01, and 701.03 (listed in order of increasing orbital period) that were reported in (*7*). Updated properties and vetting statistics for these KOIs are reported in (*8*).

After KOI 701.01-.03 were found, we found evidence of a fourth planet. We then carried out a fit of a pulse profile multiplied by a low-order polynomial at every point in the (three-planet-removed) light curve, and computed the difference in chi-square between fits with and without the pulse included. Upon inspection of the results, we found three times at which a pulse was a significantly better fit. These three locations were spaced by a separation of 267 days. We added an additional planet to our model with this ephemeris, and found an excellent fit to these three times with a transit profile of the expected duration (bottom light curve of Fig. 1). After examination of the pixel data, we confirmed that this was not due to a background body, so we promoted this to a planet candidate, KOI 701.04 (Kepler-62f after validation), with a preliminary period and epoch that were later refined via light curve modeling.

A pre-release verification and validation run of the SOC 8.3 pipeline software using Q1-Q12 data became available during the writing of this manuscript. The shallower transit of the roughly Mars-sized planet candidate with an approximate orbital period 12.44 days was identified (Fig. 1, second light curve from the top) in this run and became 701.05 (Kepler-62c, after validation).

The SOC 8.3 pipeline did not detect 701.04 because it has the minimum number of transits required for detection (three) and because the first transit occurs in the vicinity of a break in science data to downlink the data. The TPS algorithm deemphasizes cadences in the vicinity of Earth-points in a tapered fashion because residual artifacts there lead to many false positive Threshold Crossing Events (TCEs). Removal of the tapering led to a detection with an SNR of 12.

1.3 Validation tests using data characteristics

Tests on both pixel flux time series are carried out as described in (*8*) in an effort to identify astrophysical false positives masquerading as planet transits. The even-numbered transits and odd-numbered transits in the flux light curve are examined independently for each of the planet candidates. The depth of the phase-folded, even-numbered transits is compared to that of the odd-numbered transits. A statistically significant difference in the transit depths is an indication of a diluted or grazing eclipsing (or transiting) binary (or larger planetary) system. The Kepler-62 light curves show no evidence of such an odd/even effect. The difference between the ratios of the odd-numbered and even-numbered transit depths to noise for Kepler-62b through -62f are 1.5$\sigma$, 1.1$\sigma$, 2.3$\sigma$, 1.2$\sigma$, and 0.4$\sigma$, respectively.

Two methods are used to examine the motion of the image centroids to determine if they are due to a source other than Kepler-62b through -62f. The first method measures the center-of-light distribution in the photometric aperture and will be referred to as the flux-weighted centroid method. This method measures the flux-weighted centroid of every observational cadence and fits the computed transit model multiplied by a constant amplitude to the observed flux-weighted centroid motion (Table S1). The value of the constant that provides the best fit is taken to be the amplitude of the centroid motion. This amplitude is scaled by the transit depth to estimate the location of the transit source, which is used to compute the offset distance of the transit source from Kepler-62.

The second technique uses the difference-image technique and is referred to as Pixel Response Function (PRF) fitting. The PRF-fitting method fits the measured Kepler PRF to a difference image. This image is formed from the average in-transit and average out-of-transit (but near-transit) pixel images. The PRF-fitted difference-image centroid provides a direct measurement of the location of the transit signal in pixel space. This difference-image centroid position is compared with the position of the PRF-fit centroid of the average out-of-transit image, giving the offset of the transit signal source from Kepler-62.

Both centroiding methods begin in pixel coordinates. To perform multi-quarter analysis, the pixel-level results are projected onto the sky in RA and Dec coordinates. In the case of flux-weighted centroids, this projection takes place during the $\chi 2$ fit. The PRF-fitted centroids are computed quarter-by-quarter and the final results are projected into celestial coordinates. The quarterly PRF-fitted results are then averaged (minimizing a robust $\chi 2$ fit to a constant position) to account for quarterly bias due to PRF-fit error and possible crowding by nearby stars.

Both methods are subject to systematic biases (due, for example, to crowding), but the PRF-fitting method is more robust against noise and bias than the flux-weighted method. Generally the flux-weighted method's uncertainties are larger than the PRF-fit uncertainties. The quoted uncertainties do not include such systematic biases, so the measured offsets can have different statistical significance between the two methods. We therefore use both methods to increase our confidence in the accuracy of the results, though we believe the PRF-fit measurements are higher quality than the flux-weighted measurements. We only consider an offset statistically significant if it is greater than 3σ. Table S1 presents quantitative values for the observed offsets in arc seconds and in σ based on the Q1-Q12 data.

**Table S1.** Photocenter offsets with uncertainties for each planet calculated with both the PRF and flux-weighted methods.

| Planet | PRF offset (") | PRF offset (σ) | Flux-weighted offset (") | Flux-weighted offset (σ) |
|---|---|---|---|---|
| Kepler-62b | 0.24 ± 0.27 | 0.9 | 0.50 ± 0.36 | 1.4 |
| Kepler-62c | 1.34 ± 0.76 | 1.8 | 0.20 ± 1.1 | 0.2 |
| Kepler-62d | 0.16 ± 0.27 | 0.6 | 0.46 ± 0.31 | 1.5 |
| Kepler-62e | 0.14 ± 0.69 | 0.2 | 1.1 ± 0.54 | 2.1 |
| Kepler-62f | 0.81 ± 0.57 | 1.4 | 0.94 ± 0.84 | 1.1 |

The photocenter offsets are consistent with transits of the target star. Even in the worst-case uncertainty (0.72" for Kepler-62c), any object outside of a 3σ circle (2.2") can be ruled out for any of these planets. Figure S2 displays the offsets for each planet based on the PRF results and the 3σ uncertainty boundaries centered on the calculated offset for each planet. The fig. shows that the nearby star does not fall within any of the 3σ boundaries. Thus the nearby star cannot be the source of the transits.

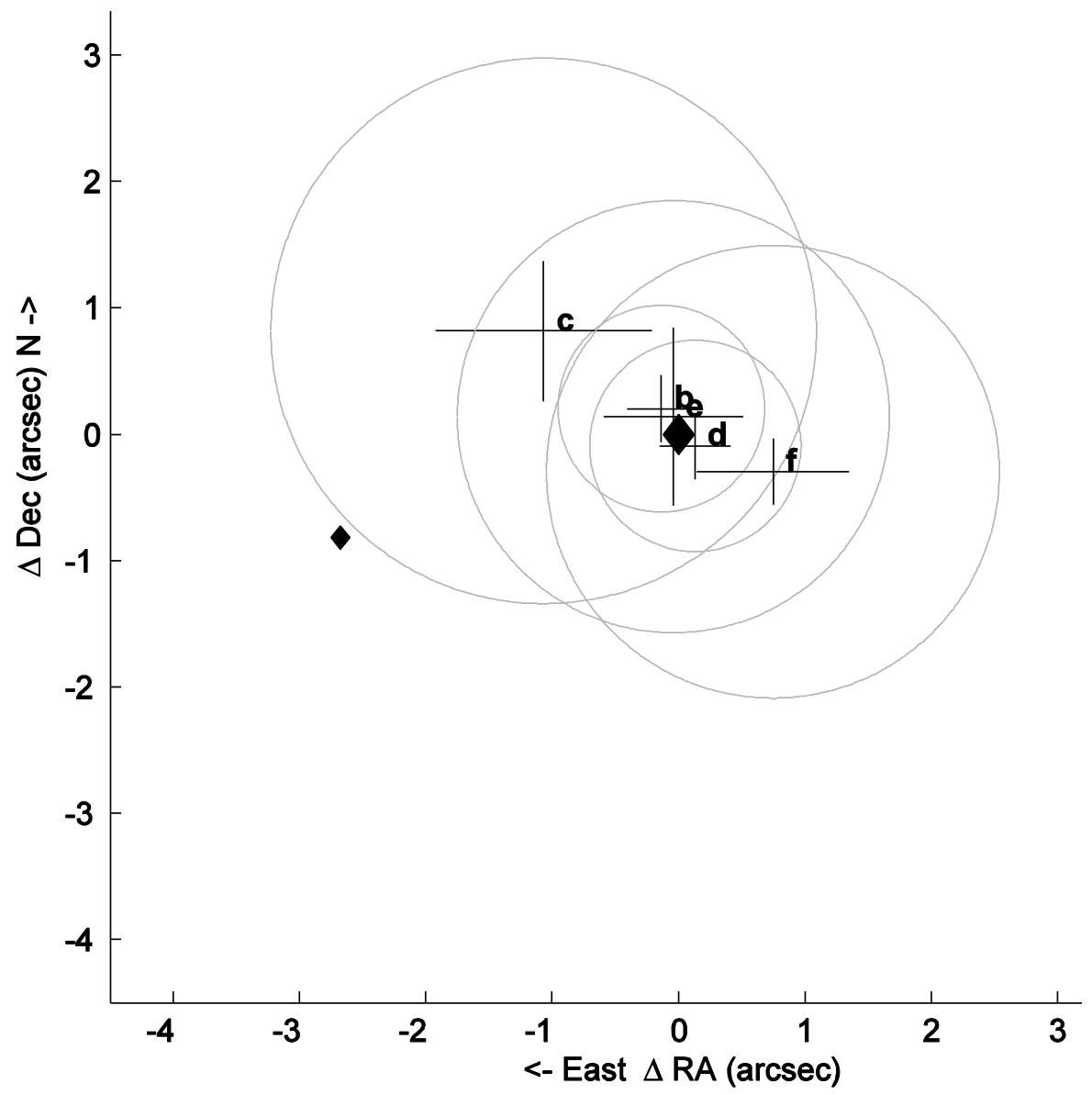

**Fig. S2.** Boundaries of uncertainty centered on the measured off-set position of the transit source for each planet. The center diamond symbol marks the known position of Kepler-62. The diamond symbol to the left marks the position of a nearby star. Each "plus" sign is the observed off-set position determined by the PRF method. The lengths of the "plus" sign indicate the uncertainty in the off-set position. The circles are centered on the observed off-set for each planet and indicate the 3σ-uncertainty boundary for each planet.

## 2. Follow-up Observations

2.1 Inspection of the nearby star field

Images were taken to inspect the star field near Kepler-62 to search for the presence of any stars that could dilute the light from the target or introduce a confounding signal. This process starts with an examination of background images ("seeing-limited images") to determine the distribution and brightness of nearby stars. Next adaptive optics (AO) are used characterize stars close to the center of the photometric aperture.

The seeing-limited image taken by the Keck 1 guider (fig. S3) shows only a single star to the East at a distance of 3" that is 6 magnitudes fainter than Kepler-62. It adds less than 1% to the visible flux in the photometric aperture.

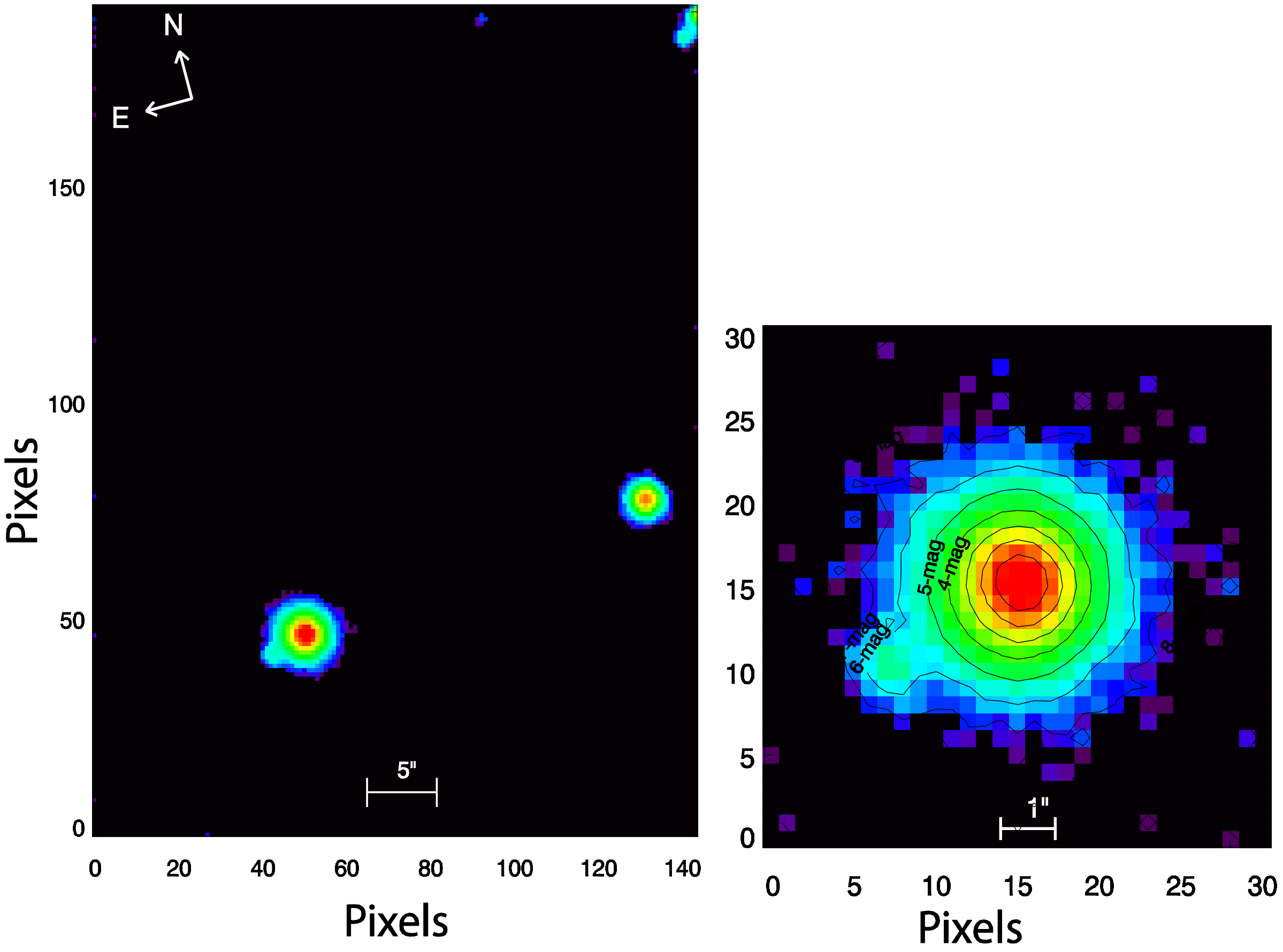

**Fig. S3.** Best-seeing images. Left panel is a 42" × 58" of the region surrounding Kepler-62 while the right panel is a 9" × 9" view centered on the target taken with the Keck 1 guider. The right panel shows the presence of a single faint star approximately 3" to the east of the target star.

Near-infrared adaptive-optics imaging of Kepler-62 was obtained on the night of 05 May 2012 with the Keck-II telescope and the NIRC2 near-infrared camera behind the natural-guide-star adaptive-optics system. NIRC2, a 1024x1024 HgCdTe infrared array, was utilized in 9.9 mas/pixel mode yielding a field of view of approximately 10". Observations were performed in the K-prime filter (K', $\lambda$ = 2.124 μm; $\Delta\lambda$ = 0.351 μm), and in the J filter ($\lambda$ = 1.248 μm; $\Delta\lambda$ = 0.163μm). Total integration times of 288 seconds and 96 seconds were taken in the K' and J filters, respectively. The frames were dark-subtracted and a flat-field correction was applied to form the final image for each filter. The central cores of the resulting point spread functions in the images have widths of full-width-half-maximum (FWHM) = 0.05" (approximately 5.7 pixels) at K' and FWHM = 0.09" (approximately 6.1 pixels) at J. The final coadded K' image is shown in fig. S4.

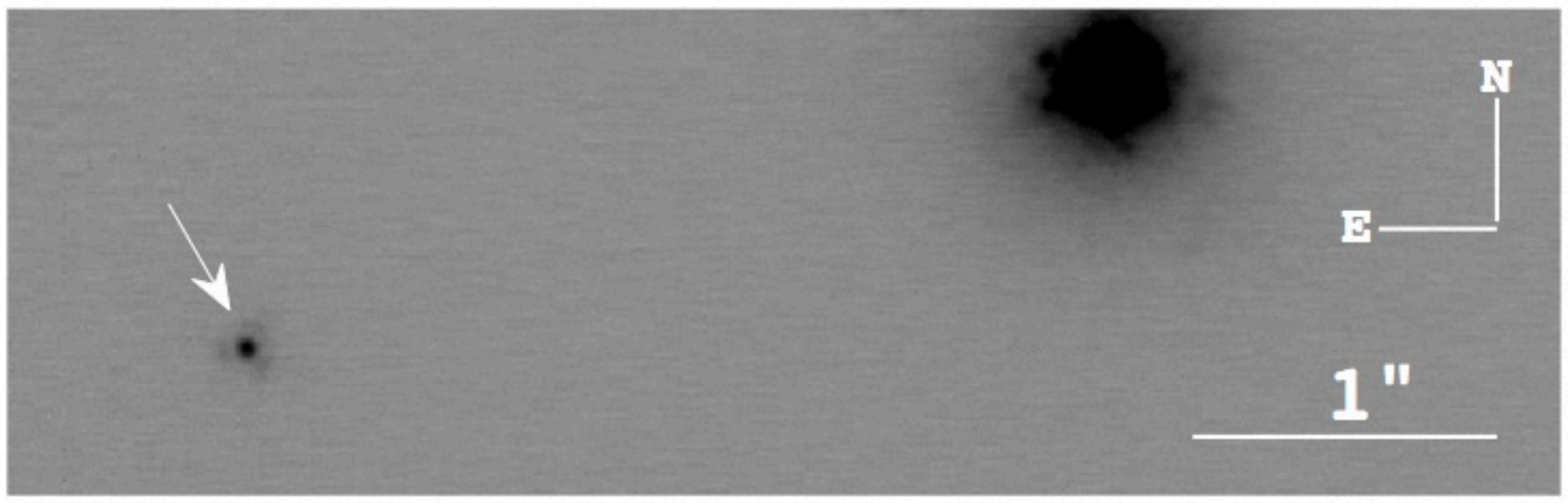

**Fig. S4.** A portion of the Keck NIRC2 K' image. The arrow points at the nearby star.

A faint source is detected 2.8" from the primary target at a position angle of PA = 107° east of north, but no other sources were detected. The source is fainter than the primary target by ΔK' = 4.12 ± 0.05 mag and ΔJ = 4.5 ± 0.1 mag, yielding magnitudes of J≈16.8 mag and K'≈15.8 mag. No other sources were detected within 5" of the primary target. The point source detection limits were estimated from a series of concentric annuli drawn around the star. The separation and widths of the annuli were set to the FWHM of the primary target psf. The standard deviation of the background counts is calculated for each annulus, and the 5σ limits are determined within each annular ring (*31*). The sensitivity curve for the K' observations is shown in fig. S5.

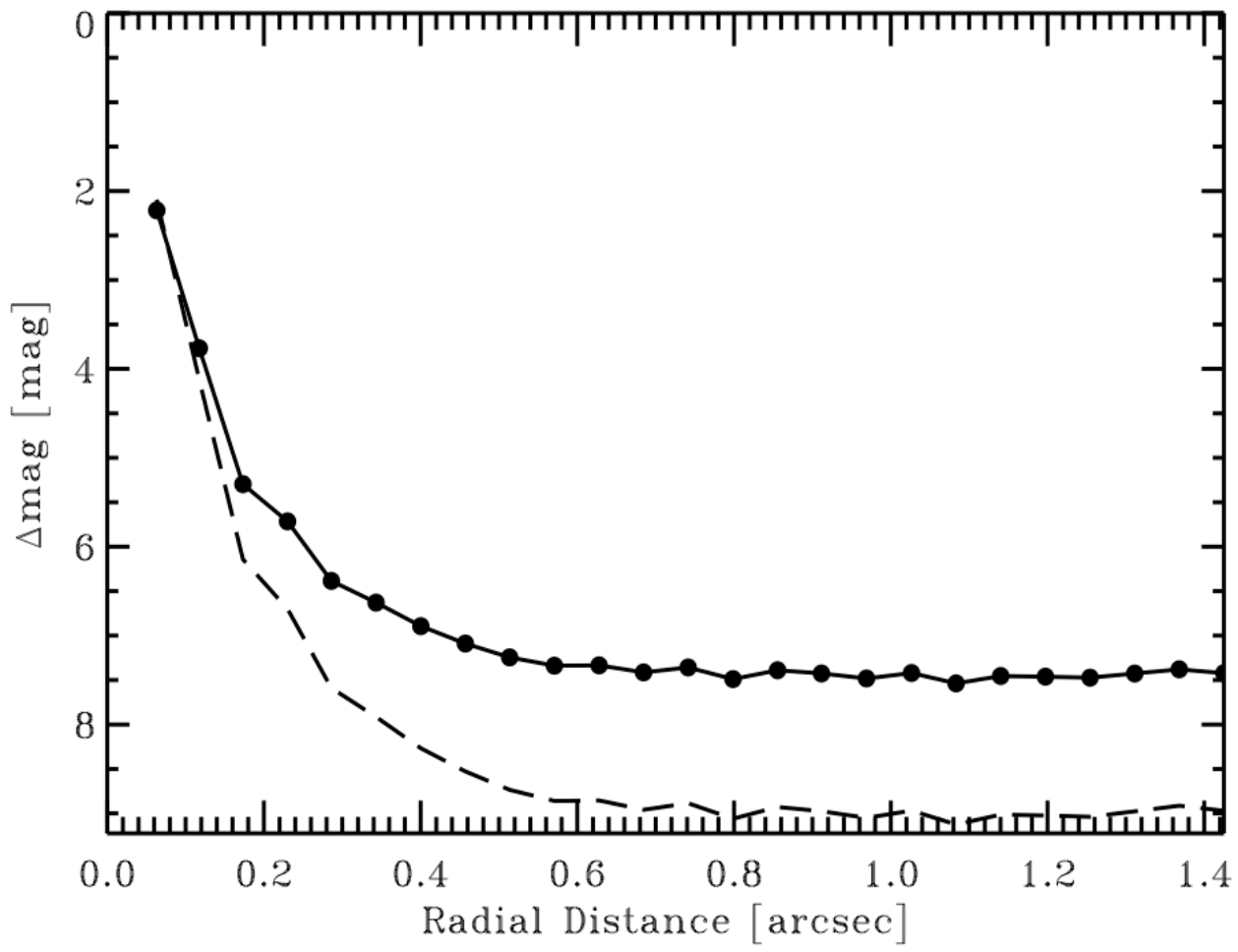

**Fig. S5.**The point source sensitivity for the Keck NIRC2 K' image, as a function of angular distance from Kepler-62. The Δmag values are 5σ limits. The filled circles represent the sensitivity limits as measured in the K' image in steps of the FWHM; the dashed lined represents the K' limits converted to estimated Kepler magnitudes based upon the measured colors of stars (*32*).

Figures S4 and S5 indicate that any background star 0.2" or further from the target would be at least 6 magnitudes fainter than the target star and thus could not be the source of the observed transit pattern. These sensitivity limits are used as input to the BLENDER analysis (described in §7) that assesses the likelihood of false-positive scenarios.

2.2 Spectroscopic check for stellar companions

Spectroscopic analysis with high-SNR data yields information about small-separation companions and complements the high spatial resolution AO imaging searches that are sensitive to separations beyond a fraction of an arc second. In searching for the second spectrum, the full range of $T_{\mathrm{eff}}$, log($g$), and [Fe/H] are examined.

The Keck HIRES spectrometer was used to take a spectrum with a resolution of 60000 and SNR = 45 per pixel in the V band for the purpose of stellar classification (see §3). The spectrum is also used to place constraints on the presence of stellar companions within approximately 0.4" of the primary star (given a 0.87" x 3" entrance slit). The analysis makes use of a library of 750 HIRES spectra of template stars spanning a wide range of effective temperatures and surface gravities. Chi-squares minimization yields the template that best matches the spectrum of Kepler-62. The best-fit library spectrum was that of HD 29883 with $T_{\mathrm{eff}}$ = 4947K, log ($g$) =4.56, and [Fe/H] = -0.15.

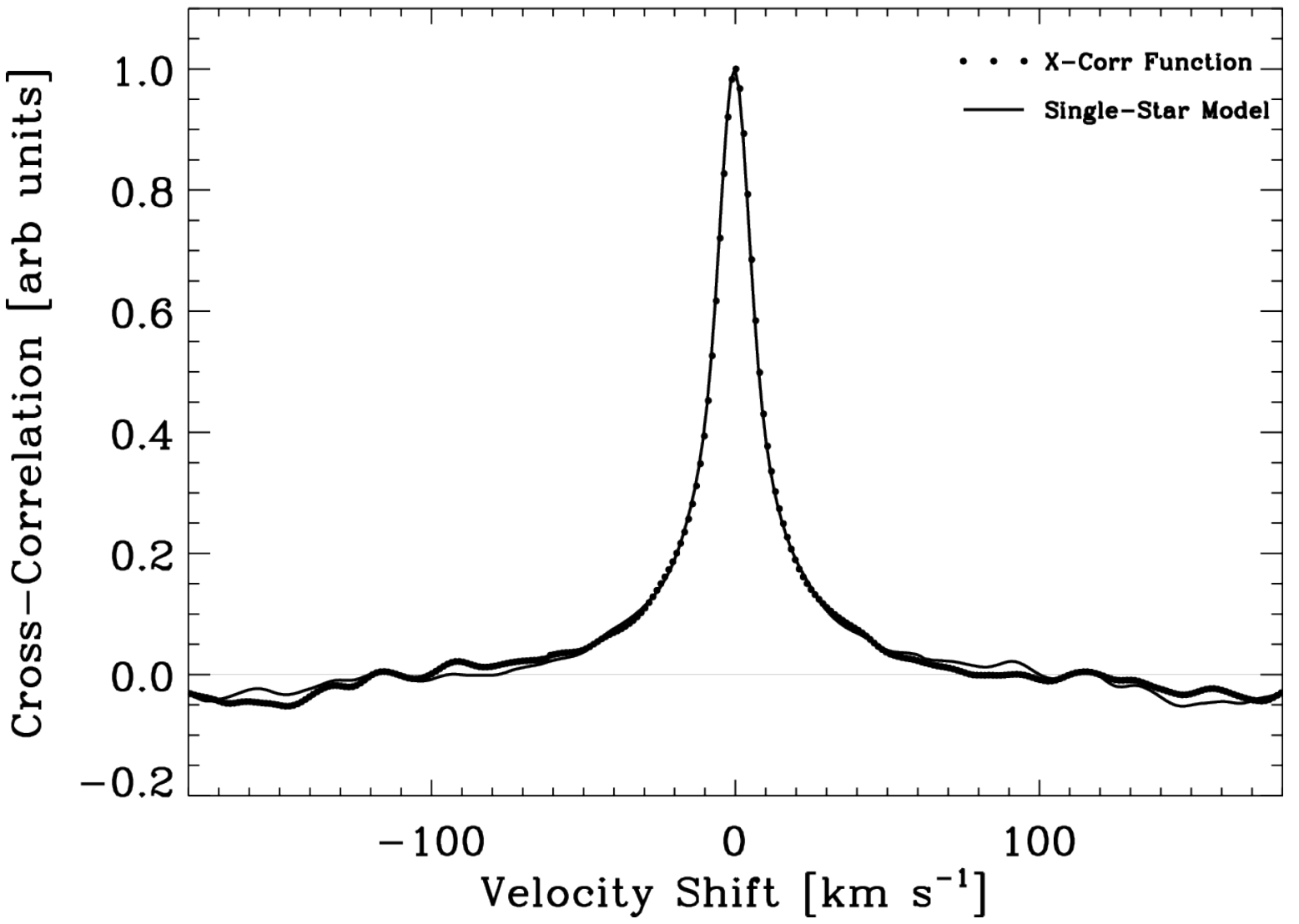

**Fig. S6**. Cross-correlation of the Kepler-62 spectrum with the best-matching template (dots) and its reflection about a vertical at the rest velocity (solid). There is no visual evidence of a second peak indicative of secondary lines indicative of a nearby star contaminating the spectrum of Kepler-62.

The cross correlation diagram presented in fig. S6 is based on a single spectrum taken without the iodine absorption cell at Keck/HIRES. By cross correlating the spectrum of Kepler-62, we are able to visualize how closely the spectrum of Kepler-62 matches the solar spectrum. The tall central peak has a 20 km/s width that is expected width for the cross-correlation function between two G-type dwarfs with absorption spectral lines having widths of roughly 10 km/s each. Discrepancies between the two are less than 3%; suggesting no secondary spectrum from a (FGK) star within 0.4” brighter than 3% of Kepler-62.

To quantify the sensitivity to the detection of secondary spectra, we proceed with a chi-square minimization against a blended spectrum: the best-matching template plus secondary of a given relative flux, spectral type and velocity offset. An example is shown in fig. S7 which plots the normalized chi-square as a function of velocity offset for the case of an M-dwarf secondary with 1%, 1.5%, and 2% the flux of the primary. Also shown for comparison are the chi-square values for the single-star case (solid line). Deviations near the rest velocity are due to slight inadequacies in the fit of the primary star.

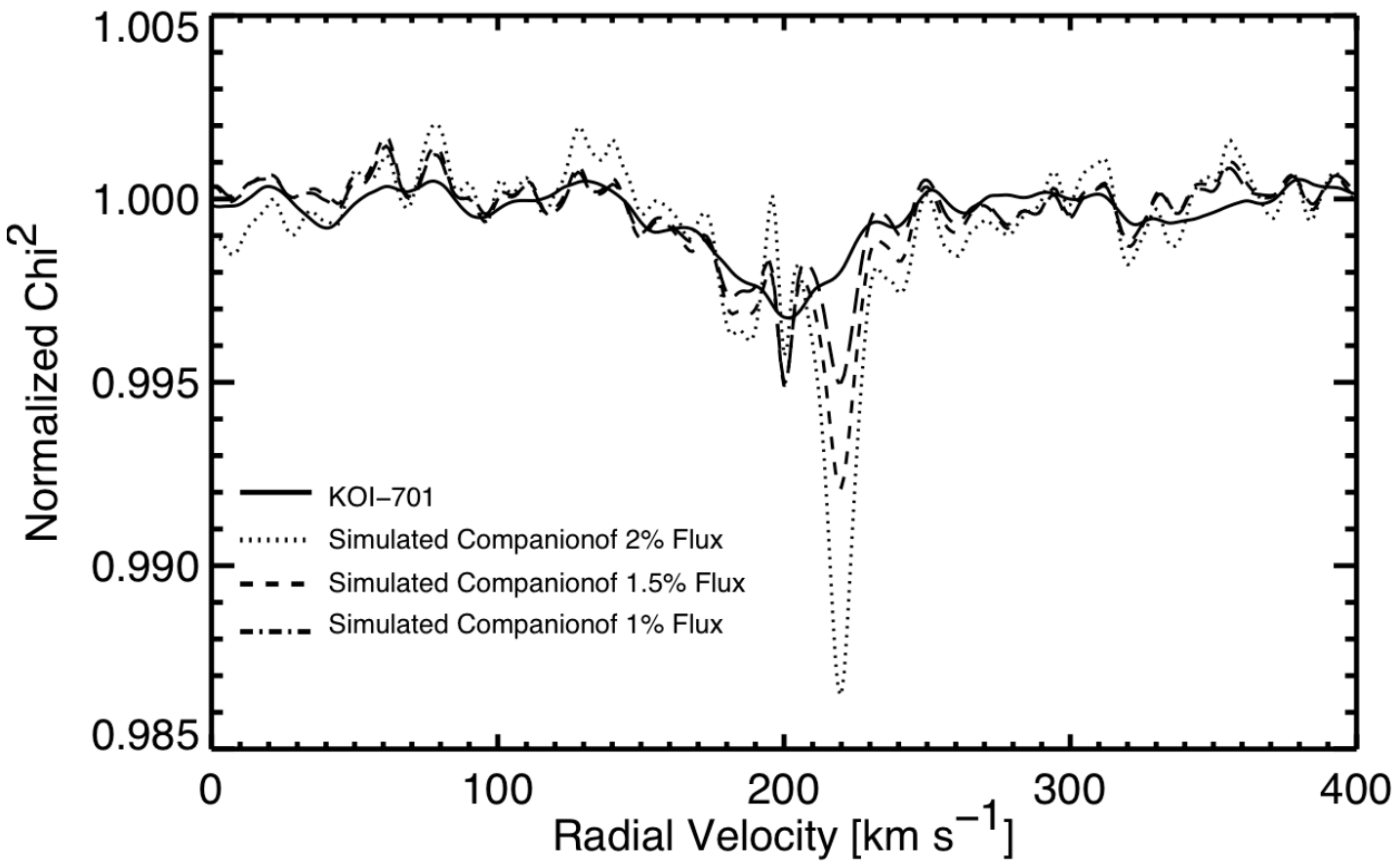

**Fig. S7**. Normalized chi-square statistic as a function of velocity offset for the Kepler-62 versus the best-matching template (solid line) and Kepler-62 versus a blend of the template plus M-dwarf spectrum at three different relative flux levels (1%, 1.5%, and 2%).

An M-dwarf having a flux 1.5% of the primary star flux (dashed black line) would stand out at the 4σ level. Companion stars as faint as 2% of the target star make a 10σ signal (dashed line). Even companions at the 1% flux level would be apparent at the 3σ level. 3σ flux limits are computed for a range of spectral types and velocity offsets. The results are shown in fig. S8.

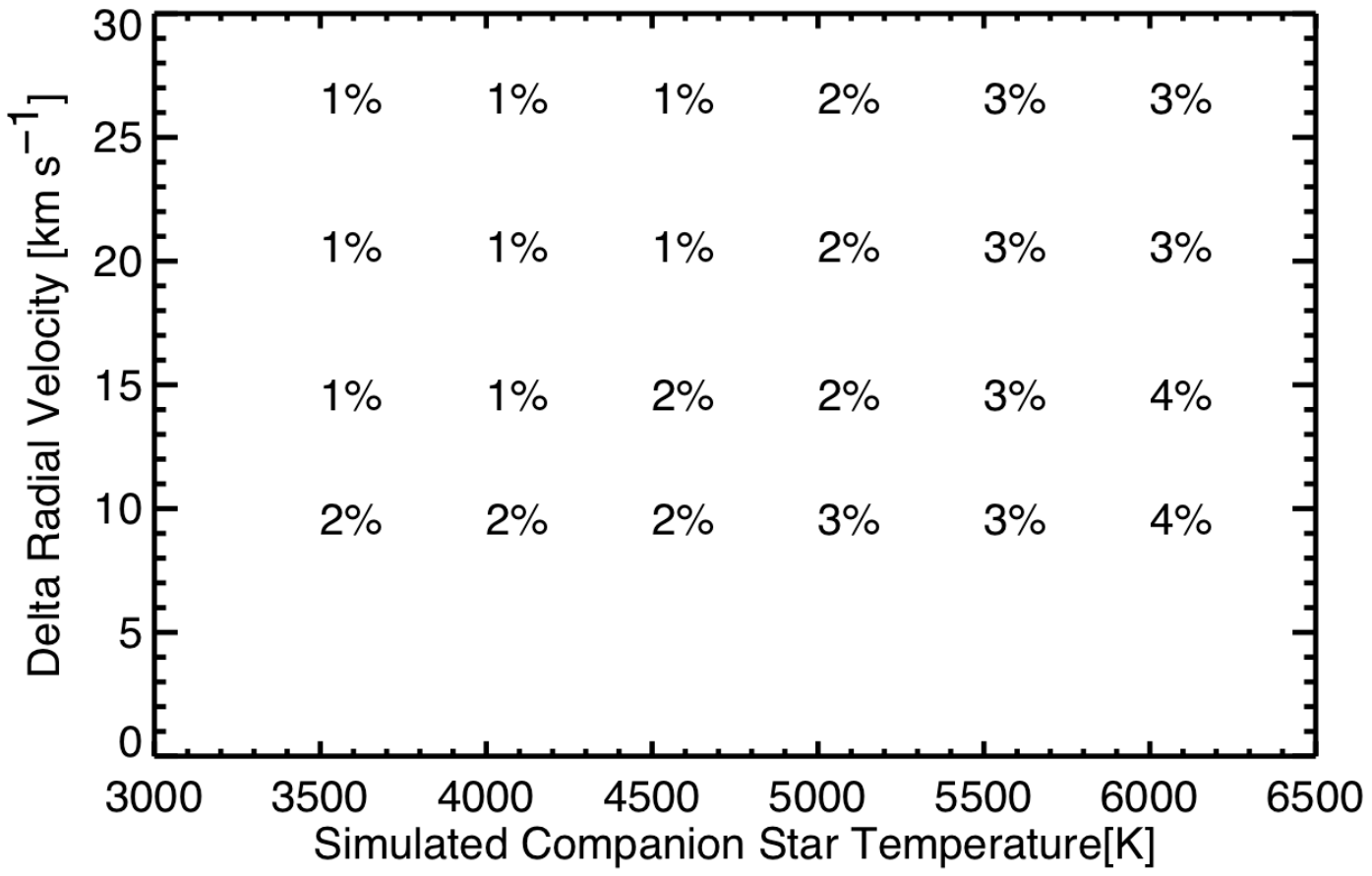

**Fig. S8.** The 3σ flux sensitivity limits as a function of spectral type and velocity offset. M-dwarf blends with just 1% the flux of the primary would be detected at the 3σ level as long as the velocity separation is not smaller than 10 km/s. The sensitivity becomes slightly worse as the spectral features of the secondary become more similar to those of the primary. There is no need to simulate relative velocities higher than 30 km/s as the spectral lines of the secondary star would be well separated from those of the primary star for all such velocities, making the secondary star equally detectable for all such velocities.

Figure S8 shows that companions of spectral type K or M would be detected down to thresholds of 1 – 2%, depending on RV separation. The entrance slit to the spectrometer is 0.87"x3" to allow us to detect any companion in that tight domain around the primary star. These detection thresholds are useful in diminishing the probability of companions, especially bound ones, around Kepler-62.

For Kepler-62 we can also rule out close binary stars based on our precise RV measurements made over a period of 128 days using the iodine cell. The dates, RV values, and uncertainties of the 13 RVs are presented in Table S2. These show a root-mean-square (rms) uncertainty of about 3 m/s, consistent with the RV errors for a g-band magnitude (similar to V) of 14.4 mag.

Any bound companion star orbiting within 10 AU would cause orbital motion in the primary star, KIC 9002278, of over 10 m/s (0.01 km/s), just due to their mutual orbital motion during 128 days. The exact RV variation of the primary star depends, of course, on the actual mass of any prospective orbiting star, its orbital phase, the inclination, and the eccentricity. But the limit of 3.5 m/s is adequate to detect a Jupiter-mass companion at 1 AU. Thus any stellar companion (with >80x the mass of a Jupiter) would make an obvious RV change, 80 times more than 3.5 m/s, if at 1 AU. Even if the stellar companion were orbiting within 10 AU, the RVs would exhibit a change of many times 3.5 m/s during the 128-day interval of the RV measurements. In summary, the analysis to explicitly search for secondary lines found none. This rules out stellar companions as faint as 1% of the flux of the primary star except for another K dwarf orbiting beyond 20 AU where the splitting the two sets of lines would not be detected.

We have also considered the possibility that a rapidly rotating G or F star could be bright enough to produce a transit of the right depth if its light is diluted by that from Kepler-62, and thus would not be readily visible in the HIRES spectrum because of its broad lines. However, nearly all G-type stars and late F-type stars rotate at about the same equatorial speed of 1 - 10 km/s. At these slow speeds, our spectroscopic method of detecting FGKM stars is effective for these types of stars, except for early F-type stars (< F2). Thus we would have detected FGKM-type stars located within 0.4" of Kepler-62, if they have a flux greater than ~2% of the primary star and have a Doppler separation of over 15 km/s. It should also be noted that rapidly rotating G or F stars are far less common than K or M stars, so that their contribution to the overall frequency of blends is negligible.

### 2.3 Warm-Spitzer results for Kepler-62e

To demonstrate the color independence of the transit depth, the transit depth of Kepler-62e was observed during one transit with Warm-Spitzer/IRAC (*33, 34*) at 4.5 μm (program ID 80117). The observation occurred on UT 2011 Oct 05. The entire visit lasted 14.42 hrs. The data were gathered in full-frame mode (256 × 256 pixels) with an exposure time of 10.4 s per image which yielded 4294 images.

A comparison of the transit depths measured by Warm-Spitzer in the infrared and by Kepler in the visible is shown in fig. S9. The measured value of $570^{+410}_{-400}$ ppm is in agreement with the Kepler measured depth at the 1σ level. If the depth was substantially larger, it would indicate the presence of a confounding star in the aperture. The agreement in the depth of the transit in both wavelength regions indicates that the ratio $R_p/R_*$ of the candidate Kepler-62e to its host star is a wavelength independent function, in agreement with that expected from a dark planetary object. Based on the assumption that all five planets orbit the same star, we determine a lower limit to the blend mass of 0.55 $M_\odot$, for Kepler-62.

**Table S2.** Keck RV observations used to estimate the upper limits to the masses of Kepler-62b, -62c, -62d, -62e, & -62f.

| Observation number | UT date | Julian date - 244,000,000 | Measured velocity (m/s) | Velocity uncertainty (m/s) |
|---|---|---|---|---|
| rj155.66 | 2012/07/29 | 16137.974661 | -4.24 | 3.78 |
| rj158.275 | 2012/09/02 | 16172.755513 | -0.80 | 3.36 |
| rj158.481 | 2012/09/03 | 16173.790612 | -5.45 | 3.27 |
| rj158.670 | 2012/09/04 | 16174.756512 | 1.72 | 3.77 |
| rj158.902 | 2012/09/05 | 16175.824954 | -3.89 | 4.32 |
| rj158.1062 | 2012/09/06 | 16176.800297 | 6.60 | 3.66 |
| rj158.1274 | 2012/09/08 | 16178.753087 | 2.08 | 3.02 |
| rj158.1383 | 2012/09/09 | 16179.793288 | -1.13 | 3.11 |
| rj159.78 | 2012/09/22 | 16192.742073 | 0.12 | 3.25 |
| rj159.687 | 2012/09/25 | 16195.812869 | -1.58 | 3.28 |
| rj161.77 | 2012/10/07 | 16207.754739 | 2.69 | 3.89 |
| rj161.489 | 2012/10/09 | 16209.792603 | 3.53 | 3.85 |
| rj163.264 | 2012/12/04 | 16265.721603 | -0.17 | 3.55 |

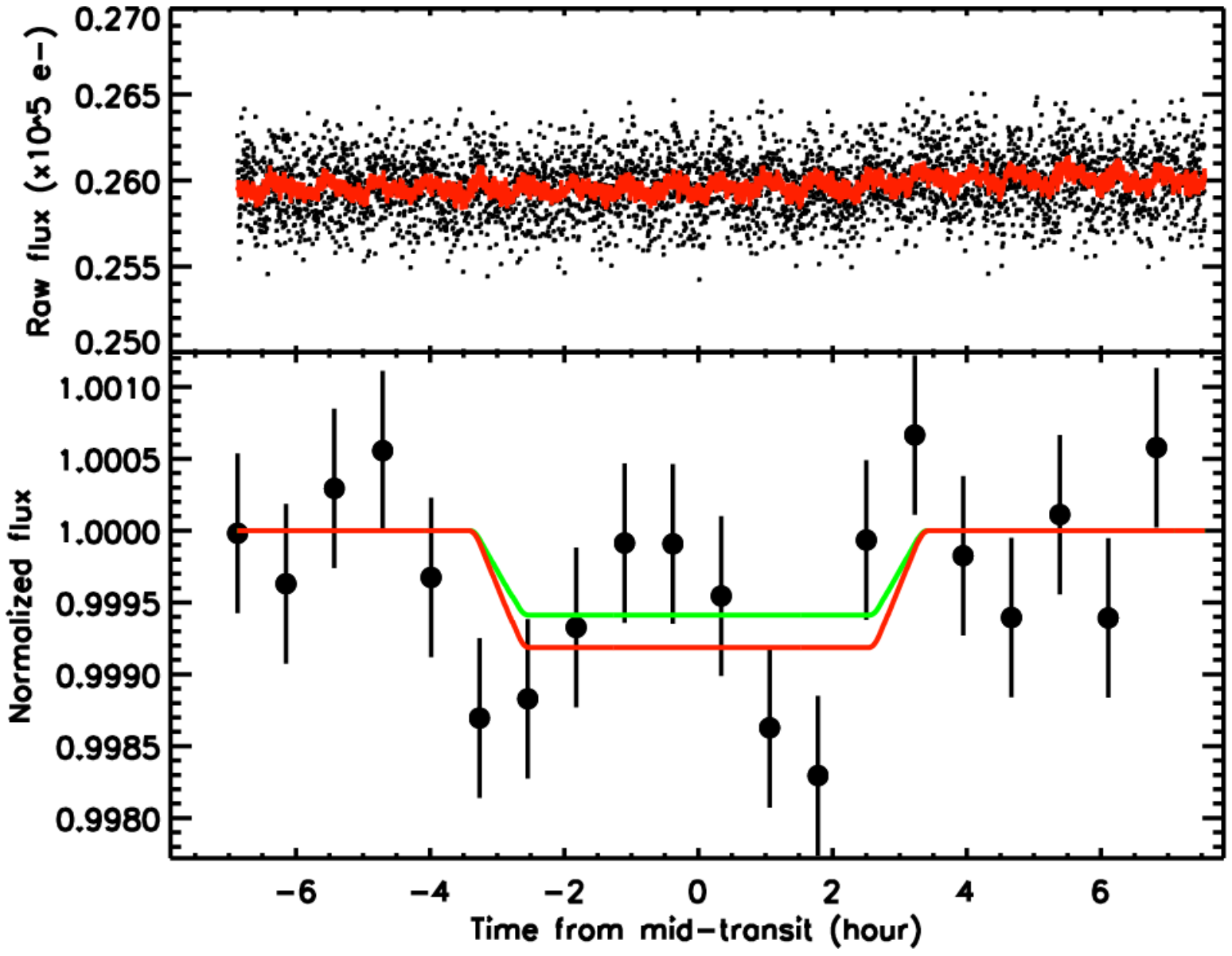

**Fig. S9.** Spitzer transit light-curve of Kepler-62e observed in the IRAC band-pass at 4.5 μm. Top panel: raw (unbinned) transit light-curve. The red solid line corresponds to the best fit model which includes the time and position instrumental decorrelations as well as the model for the planetary transit. Bottom panel: corrected, normalized and binned by 30 minutes transit light-curve with the transit best-fit plotted in red and the transit shape expected from the Kepler observations over-plotted as a green line. The two models agree at the 1σ level.

We used the APER routine to perform an aperture photometry with a circular aperture of variable radius, using radii of 1.5 to 8 pixels, in 0.5 pixel steps. A sliding median filter was used to select and trim outliers in flux and positions greater than 5σ, which correspond to 1.7% of the data. We also discarded the first half-hour of observations, which is affected by a significant telescope jitter before stabilization. The final number of photometric measurements used is 3839. The raw time series is presented in the top panel of fig. S9. Using the rms of the residual from the fit of the transit light curve, we find that the typical signal-to-noise ratio (S/N) is 150 per image which corresponds to 90% of the theoretical signal-to-noise. Therefore, the noise is dominated by Poisson photon noise.

We used a transit light curve model multiplied by instrumental decorrelation functions to measure the transit depth and its uncertainty from the Spitzer data (*13*). We compute the transit light curve with the IDL transit routine OCCULTSMALL (*35*). To obtain an estimate of the correlated and systematic errors (*36*) in our measurements, we use the residual permutation bootstrap method (*37*). We use asymmetric error bars spanning 34% of the points above and below the median of the distributions to derive the 1σ uncertainties for each parameters as described in (*38*).

## 3 Star Properties

3.1. Spectroscopic determination of host star characteristics

We obtained preliminary reconnaissance spectra of Kepler-62 with the 2.7 m Harlan J. Smith telescope at McDonald Observatory on 09:55 UT 2010 May 30, 06:21 UT 2010 July 22 and 01:39 UT 2010 October 31. These low signal/noise spectra were used to confirm the preliminary stellar properties from the Kepler Input Catalog (KIC) (*32*) and to ensure that the star was not a spectroscopic binary.

Higher SNR spectroscopic observations to obtain the best determination of the stellar characteristics of Kepler-62 were conducted at the Keck Observatory on 26 May 2011 at 13:34:39 (UT). LTE spectroscopic analysis using the spectral synthesis package SME was applied to a high resolution template spectrum from Keck-HIRES to derive an effective temperature, $T_{\text{eff}}$ = 4925 ± 70 K, surface gravity, log (*g*) = 4.683 ± 0.067 (cgs), metallicity, [Fe/H] = −0.368 ± 0.042, $v \sin i$ = 0.4 ± 0.5 km $s^{-1}$, and the associated error distribution for each of them.

3.2. Fundamental stellar properties

As described in §3, spectroscopic observations were used to derive the stellar effective temperature, log(*g*), and metallicity. By matching these values to Yonsie-Yale stellar evolution models (*39, 40*), the stellar size, mass, luminosity, and age were estimated. The model matching was done by varying the stellar mass, age and [Fe/H] and comparing the model-derived values of $T_{\text{eff}}$, log(*g*) and [Fe/H] to the spectroscopic values with a chi-square statistic. An initial match was found by scanning in mass increments of 0.1 $M_\odot$ and restricting ages from 0 to 14 Gyr and identifying a best matching model (fig. S10). A Markov-Chain-Monte-Carlo (MCMC) routine was then seeded with this trial value of stellar mass, age and [Fe/H] to determine posterior distributions. In total 100,000 chain elements were generated. The model was also used to determine posterior distributions for the stellar radius, luminosity and mean stellar density. For each stellar parameter we report the median and standard deviations. These are listed in **Table S3.**

**Table S3.** Adopted stellar parameters for Kepler-62.

| Parameter | Adopted Value | Notes |
|---|---|---|
| Right Ascension (J2000) | $18^h 52^m 51.06^s$ | A |
| Declination (J2000) | $+45^\circ 20' 59.50''$ | A |
| Kepler Magnitude | 13.75 | A |
| R magnitude | 13.65 | A |
| Effective temperature $T_{\text{eff}}$ (K) | 4925 ± 70 | B |
| Metallicity [Fe/H] | -0.37 ± 0.04 | B |
| Gravity log(*g*) (cgs) | 4.68 ± 0.04 | B |
| Projected rotation velocity $v \sin i$ (km/s) | 0.4 ± 0.5 | B |
| Mass $M_*$ ($M_\odot$) | 0.69 ± 0.02 | C |
| Radius $R_*$ ($R_\odot$) | 0.64 ± 0.02 | C |
| Density $\rho_*$ (cgs) | 3.8 ± 0.3 | C |
| Age (Gyr) | 7 ± 4 | C |
| Luminosity $L_*$ ($L_\odot$) | 0.21 ± 0.02 | D |
| Rotation Period (days) | 39.3±0.6 | E |
| Distance (pc) | 368 | F |

Notes: A; from KIC (*32*), B; from analysis of high resolution spectra, C; modeling results based on

measured values of $T_{eff}$ and gravity plus Yonsei-Yale evolutionary curves (*39, 40*), D; calculated from stellar size and $T_{eff}$; E; from Fourier analysis of photometric light curve, F; determined from derived Rmag, conversion to Vmag equals 0.73, interstellar extinction = 0.15magV, and for an absolute visual magnitude of 6.4 for a K2 dwarf.

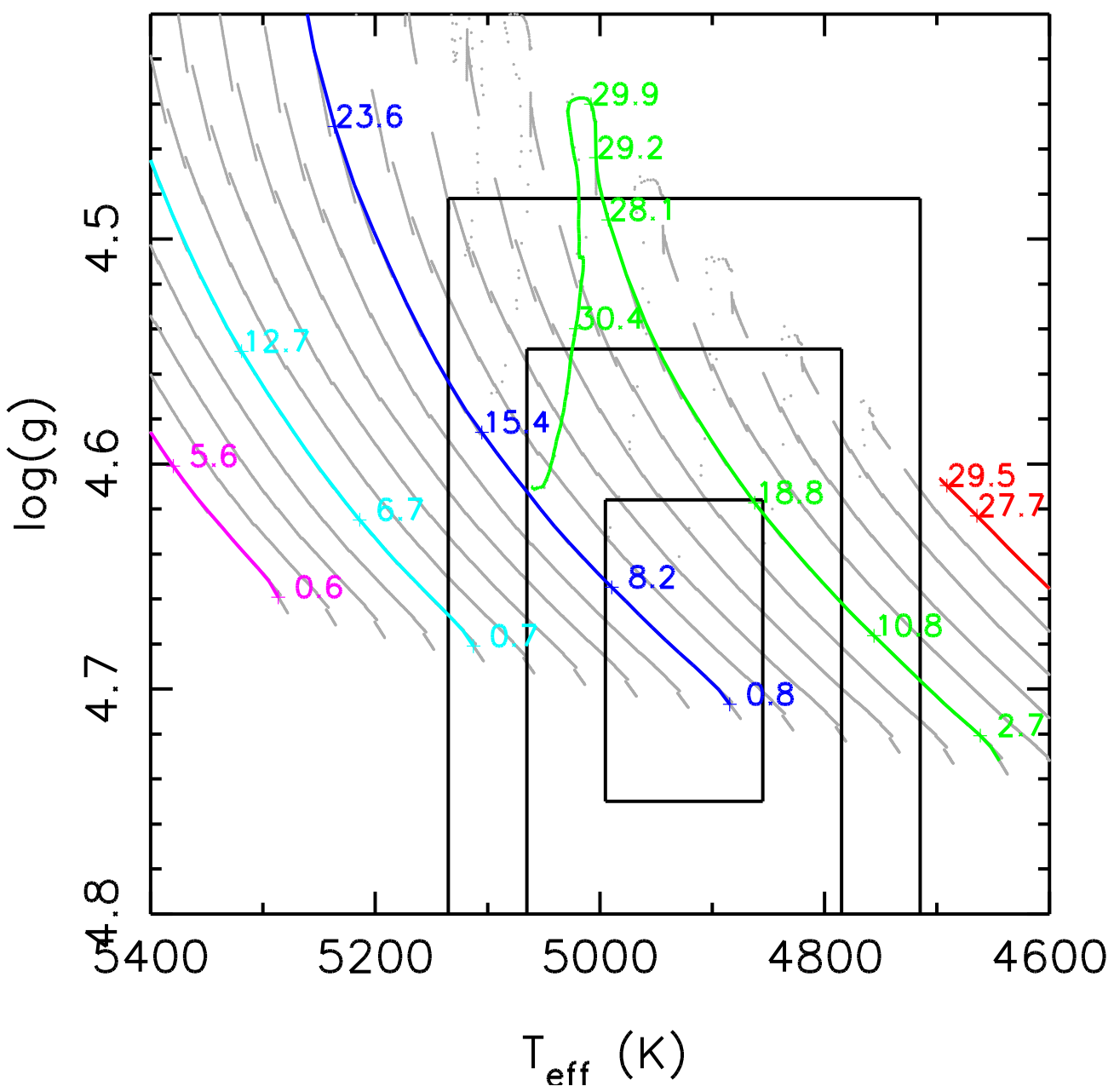

**Fig. S10.** Yonsei-Yale evolutionary curves based on the measured log(*g*) and $T_{eff}$ were used to deduce stellar mass, density, age and their uncertainties. The three boxes show the 1, 2, and 3σ uncertainties from SME spectroscopic analysis. The tracks show the evolution-model grid starting at 0.6 $M_\odot$ on the right to 0.79 $M_\odot$ on the left in 0.1 $M_\odot$ intervals. The red, green, blue, cyan and magenta lines highlight the 0.6, 0.65, 0.7, 0.75 and 0.79 $M_\odot$ tracks. The labels indicate the model ages in Gyr.

### 3.3. Activity and rotation

Figure S11 shows the power spectrum of the Kepler-62 light curve for quarters Q1-Q12. A periodogram analysis reveals a significant peak at 39.3 ± 0.6 days. Assuming this corresponds to the stellar rotation period, the age-rotation relation (*41*) can be used to estimate the age of the star. We find an estimated age of 6.5 ± 0.2 Gyr, in agreement with the value determined from isochrone fits to the spectroscopic $T_{eff}$, log(*g*), and [Fe/H] values.

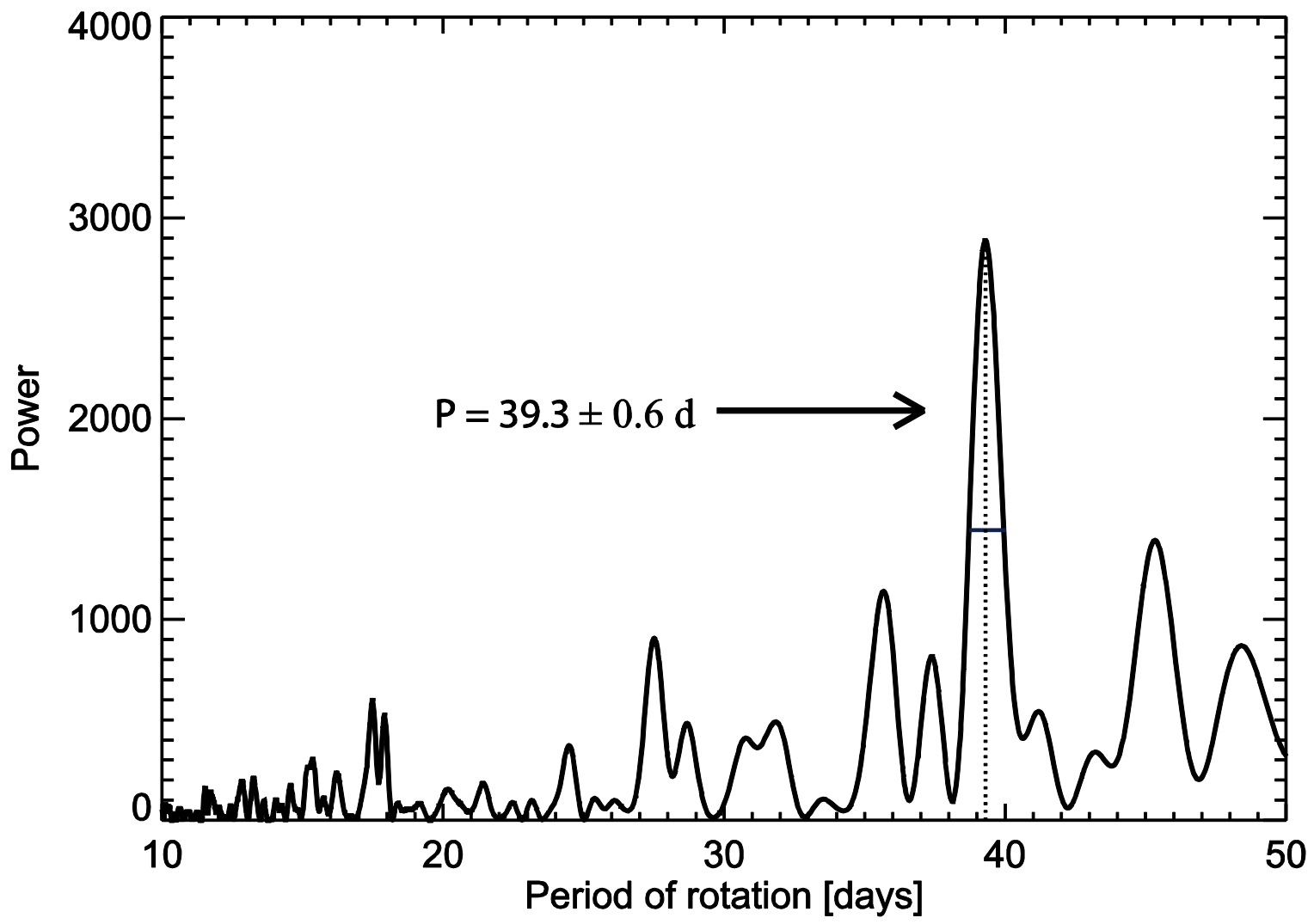

**Fig. S11.** Power spectrum of Kepler-62 showing a strong peak at 39.3 days.

It is also possible to estimate the age and rotation rate using the CaII H &K emission index log(R'$_{HK}$). Following the methodology of (*42*), we simultaneously measure the CaII H&K lines of Kepler-62 with each radial velocity observation. (These lines are often used tracers of stellar chromospheric activity.) As detailed therein, we transform these measurements into the S index of chromospheric activity (*43*) finding S = 0.212 ± 0.002. Given the empirical relationship between chromospheric activity and rotation period, and hence age (*44, 45*), we use our CaII H&K measurements to estimate the age of Kepler-62 with the gyrochronology relations commonly used to estimate the ages of field stars (*46, 47*). Taking the B-V color of 0.832, as measured by (*48*) and following the prescription of (*45*), we transform our S index into the log(R'$_{HK}$) index of activity, related to the former but corrected for photospheric contributions. We find log(R'$_{HK}$) = -4.863 ± 0.006, an inactive star according to (*49*). (see also (*42, 50*). With this value and using the relations of (*47*), we derive a rotation period of 37 ± 6 days, in agreement with the photometrically measured period of 39.3 ± 0.6 days, and we estimate a stellar age of 5.4 ± 1.7 Gyr, comparable to the age of 6.8 Gyr obtained from stellar evolutionary models.

## 4. Search for Transit Timing Variations (TTV)

We fit transit models (*35*) to the data, short cadence where available and long cadence elsewhere, with mid-transit time being a free parameter. We found no convincing transit timing variations, and therefore calculate only upper limits to the planet masses from these data. For planets -62e and -62f, we calculated the standard deviation of the offset of the times from a constant period, after resampling them according to their error bars, and are able to limit at a 95% confidence level, the standard deviation to less than 20 minutes and 14 minutes for planets -62e and -62f respectively. Planet -62f has only 3 transits measured so far, and the third transit falls within ~ 5 minutes of its expected position given the period established by the first two transits. In principle, even a large TTV signal might have 2 consecutive periods that are the same length. Therefore we asked not whether certain masses could match that constraint, but rather how fine-tuned such a situation is. We turned to numerical integrations to determine this, as follows.

We picked osculating Jacobian periods and transit epochs close to the observed values, and sampled a grid of ecosω and esinω values for each planet, between ± 0.3 with steps of 0.03, resulting in 194,481 simulations. We pared down the simulated transits to only those that were observed, and computed the standard deviation of their TTV about their own best-fitting linear ephemeris. The dynamical effects the data can probe at this point are all in the linear regime, i.e., the amplitude of the signal scales with the mass of the perturbing planet (51). Therefore we picked 10 $M_\oplus$ for both planets in these investigations, and scaled the signal to make inferences about other masses.

At particular masses, larger eccentricities give larger TTV signals (*52*). At the masses used for the simulation (10 $M_\oplus$ each), eccentricities of ~0.3 gave TTV signals comparable to the limits. At low eccentricities ≤ 0.1, masses above 150 $M_\oplus$ for planet e and above 35 $M_\oplus$ for planet -62f contradicted the timing data in more than 95% of trials. Therefore our 95% confidence limits on the masses of planets -62c, -62d, -62e, and -62f are 4, 120, 150, 35 $M_\oplus$, respectively. In all of these cases, the transit times do not yield physically interesting constraints on the densities of the planets, which may be up to about 100 g $cm^{-3}$ and still agree with the timing data.

In addition, we also carried out a separate analysis in which we computed a dynamical fit to the 11 transit times of planets -62e and -62f through Q13. The planets were assumed to be coplanar, giving 10 model parameters (planet/star mass ratio and four orbital elements for each planet), leaving two degrees of freedom. Over a grid of mass ratios for the two planets, we held the mass ratio of each planet fixed, and minimized the chi-square of the fit with respect to all of the other parameters in the model (requiring Hill stability for each of the computed models). The fits result in an upper limit of 51.5 $M_\oplus$ ($3\sigma$) for Kepler-62f, while no upper limit on Kepler-62e was found up to the largest mass ratio of 0.003. These two analyses give consistent results for Kepler-62f, and are compatible for Kepler-62e considering that the second analysis did not penalize large eccentricities.

**5. Mass Constraints from High-Precision Doppler Measurements**

We obtained 13 Keck-HIRES RVs for Kepler-62 spanning 128 days in 2012, from BJD=2456137.97 to 2456265.72. See Table S2 for the list of all times, RVs, and uncertainties which include jitter of 2 m/s added in quadrature. With typical exposure times of 45 minutes, we achieved a SNR of 70 per pixel.

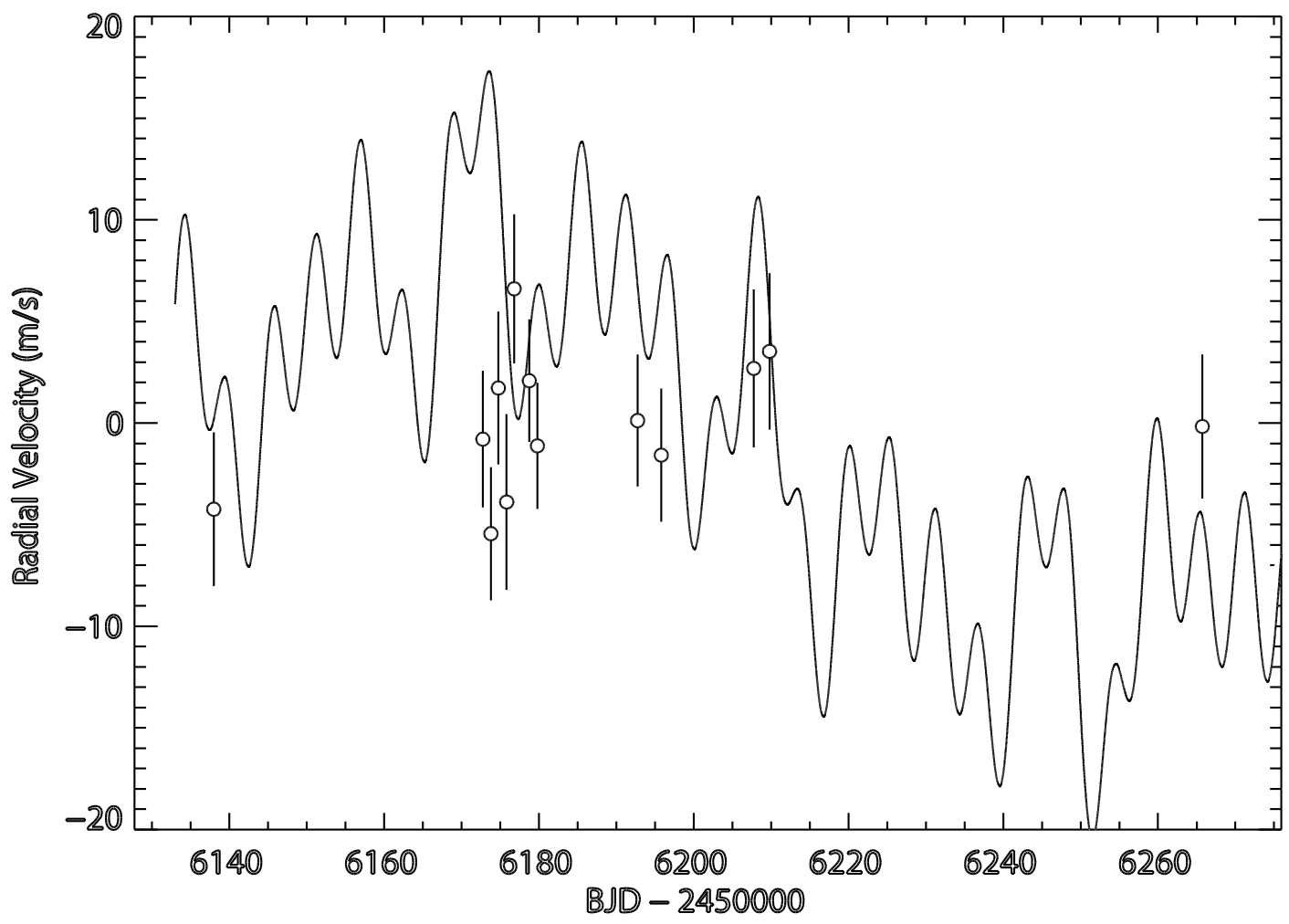

**Figure S12**. Upper limit to the sum of the RV amplitudes (solid curve) versus time based on the upper limits to the RV for each of the five planets. Measurements are shown as dots with 95th percentile error bars.

The upper limit to the radial velocity of each planet is found that would be inconsistent with the measurements. Given the periods and ephemerides for all five planets, an MCMC analysis provides the posterior distribution of the planet masses and densities. Upper mass and density limits (95th percentile of the posterior distribution) for all five of the transiting planets were calculated and are tabulated in Table S4.

Figure S12 compares the sum of the calculated upper limits with the RV measurements (Table S2). The results indicate that the measured RV values are consistent with planetary mass objects, but do not yield physically interesting constraints on the planet densities.

**Table S4.** Upper limits to the masses and densities of Kepler-62b through -62f

| Planet | Period (days) | TTV Results (95th percentile) | | RV Results (95th percentile) | |
|---|---|---|---|---|---|
| | | Upper limit to the mass ($M_\oplus$) | Upper limit to density (g/cc) | Upper limit to the mass ($M_\oplus$) | Upper limit to density (g/cc) |
| Kepler-62b | 5.7 | | | 9 | 22 |
| Kepler-62c | 12.4 | 4 | 140 | 9 | 338 |
| Kepler-62d | 18. | 120 | 90 | 14 | 11 |
| Kepler-62e | 122 | 150 | 200 | 36 | 47 |
| Kepler-62f | 267 | 35 | 70 | 43 | 85 |

**6. Planet Properties from Model Results**

To determine the planetary parameters for Kepler-62b though -62f, we started with the Q1-Q12 Simple Aperture Photometry long-cadence light curve. The time series was then detrended with a running 2 day median filter using the method described in (*18*). Observations that occur during a planetary transit event were excluded from the calculation of the median and each quarterly light curve was normalized by the quarterly median prior to detrending. The resultant flux series (bottom panel of fig. S1) was used to model the planetary transits.

The photometric model assumes non-interactive Keplerian orbits and uses the quadratic transit model of (*35*). We used the limb-darkening parameters from (*53*) which were fixed to 0.5396 and 0.1731. The model is parameterized by the mean-stellar density ($\rho_*$), photometric zero point epoch ($T_0$), period ($P$), scaled planetary radius ($R_p/R_*$), impact parameter ($b$) and eccentricity ($e$) and the argument of periapse ($\omega$) parameterized as $e\sin\omega$ and $e\cos\omega$. The semi-major axis for each planet is estimated by $(a/R_*)^3 \sim \rho_* * G * P^{2/3}$, where $G$ is the gravitational constant and the assumption is made that the sum of the planetary masses is much less than the mass of star. For a Jupiter-mass companion, a systematic error of 0.02% would be incurred for the measurement of $\rho_*$.

A best fit model was calculating by a Levenberg-Marquardt chi-square minimization routine. In this model, the mean stellar density was fixed to the stellar model value from Table S3. The best-fit model was then used to seed a MCMC routine (18) to determine *posterior* distributions of all the model parameters. We used the *posterior* distribution for the mean-stellar density as determined in Table S3 as a constraint. This restricts the allowed solution space of model parameters that are correlated with the stellar density $\rho_*$ such as the impact parameter $b$ and $e\sin\omega$.

We ran the MCMC algorithm 4 separate times, each time generating 1,000,000 elements. The first 10% of each chain was discarded and then the four chains were combined and used to produce *posterior* distributions for each parameter. We report the median value and ± 68 percentiles for each parameter in Table 1. It should be noted, that using the constraint on the mean-stellar density drives the model parameters of planet -62e towards an eccentric orbit. This informs us that the stellar parameters are not consistent with a model that assumes a circular orbit. A circular model would require a mean-stellar density of 2.7 g $cm^{-3}$ which disagrees with the stellar models. Also, such a low stellar density is incompatible with stellar evolution theory for a star with $T_{\rm eff}$ = 4925 K that has an age less than 14 Gyr.

We also tested the long-term stability of the system based on the parameters reported in Table 1, with nominal masses based on a fit to the planet mass-radius relation of solar system planets, $M_p/M_\oplus = (R_p/R_\oplus)^{2.06}$ (*54*). While planets -62c and -62d quickly collided when we used the values of ecosω and esinω from Table 1, we found no signs of long-term instability when starting from nearly circular orbits based on 10 Myr integrations. The system could also be rendered stable by reducing the masses of planets -62c and/or -62d relative to the above mass-radius relation or by a combination of reduced masses and eccentricities.

## 7. Validation

Dynamical confirmation of the planetary nature of the signals presented by Kepler-62 has traditionally required either a detection of the reflex motion of the star (Doppler signature) or a detection of transit timing variations from the mutual interactions among the objects. In the absence of such tell-tale diagnostics, we take an alternative approach. We perform follow-up observations designed to that rule out as many astrophysical false positive scenarios as possible and then estimate, via numerical simulations, the probability that the signal is due to one of the remaining false-positive scenarios that could not be ruled out by observations. We compare this probability to the likelihood of the planet interpretation and consider a planet to be statistically validated if this odds ratio exceeds 400:1 (approximately corresponding to the cumulative distribution function at 3-sigma for a normal distribution). The numerical simulations are part of the BLENDER procedure (*10-13*) and used to validate a number of other Kepler planets (*14-18*). We refer the reader to these works for the full details and offer only a brief summary here.

We performed a systematic exploration of the different types of false positives that can mimic the signals by generating large numbers of synthetic blend light curves over a wide range of parameters and comparing each of them with the Kepler photometry in a chi-square sense. We rejected blends that result in light curves inconsistent with the observations. We then estimated the frequency of the allowed blends by taking into account all available observational constraints from the follow-up observations mentioned above. Finally we compared this frequency with the expected frequency of true planets (planet "prior") to derive the "odds ratio".

The types of false positives we considered include eclipsing systems falling within the Kepler aperture that are either in the background or foreground, or that are physically associated with the target. We allowed the object producing the eclipses to be a star or a planet. The observational constraints we used include the following: (a) the color of the star as reported in the KIC which allows us to rule out any simulated blends resulting in a combined-color that is significantly redder or bluer than the target; (b) limits from the centroid motion analysis on the angular separation of companions that could produce the signal (Sect. 1); (c) brightness and angular separation limits from high-resolution adaptive optics (Sect. 2.1); (d) limits on the brightness of unresolved companions from high-resolution spectroscopy (Sect. 2.2); and (e) a constraint from the measured transit depth of Kepler-62e derived from our Spitzer observations (Sect. 2.3), that place an upper limit on the mass (spectral type) of stars producing the blend. For eclipsing systems physically associated with the target, we also considered dynamical stability constraints in hierarchical triple configurations (*55*). To estimate the planet prior we relied on the list of candidate planets (KOIs) by (*8*), restricted to main-sequence host stars and with the assumption that the list is complete (i.e., that all signals have been detected) and that the rate of false positives is negligible. Our simulations for the signals discussed here indicate that the contribution of background eclipsing binaries to the blend frequencies is nearly insignificant in all cases. Background stars transited by large planets, on the other hand, can more easily mimic the signals, as can stars that are physically associated with the target and that are transited by a large planet.

For Kepler-62b (KOI-701.02) we found that the frequency of background/foreground blends is $6.04 \times 10^{-8}$, while that of blends involving larger planets transiting physical companions to the target is $1.02 \times 10^{-7}$. The planet prior was estimated by counting the number of known KOIs (61 in this case) that are in the same radius range (within $3\sigma$) and period range (within a factor of 2) as the putative planet. The same period constraint is used for the blend population. We account for limitations in the sample completeness and reliability following the procedures described in (*56*).

These simulations suggest that 5.8 of the 61 KOIs may be false positives and that a completeness factor of approximately 2.19 is required. That is, a signal like that of Kepler-62b could have been detected around only 46% of the main sequence Kepler targets. The corrected planet count is then (61 - 5.8)*2.19 = 120.9. With this, the planet prior becomes 120.9 / 138,253 = $8.74 \times 10^{-4}$. The final odds ratio for Kepler-62b is then $8.74 \times 10^{-4}$ / ($6.04 \times 10^{-8}$ + $1.02 \times 10^{-7}$) = 5400 which allows us to validate the planet with a very high degree of confidence. We note that the issues of completeness and reliability of the KOI catalog apply to both the planet prior and the blend frequencies since the latter draws upon the catalog to inform the rate of occurrence of large transiting planets comprising the blend.

Applying a similar procedure to Kepler-62d (KOI-701.01), we found a background blend frequency of $2.06 \times 10^{-9}$, and a frequency of blends involving physically associated companions of $5.53 \times 10^{-8}$. For the planet prior we tallied 88 KOIs in the relevant radius and period range, of which we expect 10.9 may be false positives. The incompleteness boost is a factor of 1.54 in this case. The planet prior is then (88 - 10.9)*1.54 / 138,253 = $8.59 \times 10^{-4}$, and the odds ratio becomes approximately 15,000. This is also high enough to clearly validate the signal as being of planetary nature.

For Kepler-62e (KOI-701.03), which is the signal corresponding to a super-Earth-size planet in the habitable zone, the background blend frequency is $2.81 \times 10^{-9}$ and the frequency of blends involving physically associated companions is $3.48 \times 10^{-9}$. Because very few candidates like this have been found, the incompleteness correction is larger than for the other two signals, and comes to a factor of 4.08 according to our simulations. The false positive rate is also more important. The planet prior for this case is (4 - 0.87)*4.08 / 138,253 = $9.24 \times 10^{-5}$, and the odds ratio becomes $9.24 \times 10^{-5}$ / ($2.81 \times 10^{-9}$ + $3.48 \times 10^{-9}$) = 14,700. This again validates the signal to a high degree of confidence.

The unexpectedly large odds ratio for Kepler-62e, a small planet with a very long orbital period, is due to the lower incidence of blends involving physically associated companions, which dominate the total blend frequency for the other two signals. The reason these types of blends are less frequent for Kepler-62e has to do with the long duration of the transit. In order for a planet orbiting a companion star to reproduce this long transit duration its orbital speed must be slower than in a circular orbit of the same period, which generally requires eccentric orbits and transits occurring near apastron. As it turns out, most of the low and modest eccentricity cases allowed by BLENDER are ruled out by other observational constraints (color information, spectroscopic limits), and only the most eccentric cases ($e > 0.76$) with apocentric transits remain viable. Those scenarios, however, are very unlikely.

BLENDER constraints on false positives for each of the five planets can be seen in Fig.2, for each of the three blend scenarios found to be relevant here (background eclipsing binaries, background/foreground stars transited by a planet, and physically associated stars transited by a planet). Also shown are the complementary constraints afforded by the follow-up observations: cross-hatched areas for regions excluded by the color information (cyan) or spectroscopic limits on the brightness of unresolved companions (green), and the gray area in the panel for Kepler-62e for the limits from our Spitzer observations on the mass of potential blended stars. Each panel shows a cross-section of the space of parameters for blends, and a contour enclosing the area in which the light curves produced by false positives yield an acceptable fit to the Kepler photometry (within $3\sigma$ of the best-fit transit model). Only blends within these regions count towards the blend frequencies given above. Blend scenarios outside of this area are ruled out, as are blends that are covered by the cross-hatched or gray regions.

The validation of Kepler-62c (KOI-701.05) and Kepler-62f (KOI-701.04) proceeds exactly as for Kepler-62b, -62d, and -62e with one exception. The long period of Kepler-62f and the small size of -62c make these standouts in the KOI catalog. They are practically the only objects of their kind detected to date: there is only one other KOI within each of the radius/period bins defined by Kepler-62c and -62f. Consequently, the construction of a reasonable planet prior relies on extrapolation from other regions of parameter space. The situation is especially problematic for Kepler-62f since a significant source of astrophysical false positives is a background blend with a transiting Neptune-size planet. The sample of Neptune-size planets out at periods >200 days is also too small to reliably inform the numerical simulations.

For Kepler-62c, we assume that the planet occurrence rate as a function of size is flat for short-period (6 to 24 days) planets smaller than $1R_{\oplus}$. We follow the procedures outlined above and then ask the question: how much smaller would the true occurrence rates have to be to yield a validation at exactly the 99.7% confidence level? We begin with the occurrence rates presented in (*56*) that are reproduced in fig. S13 (solid black line) and extrapolate to smaller sizes by using the average of the two smallest-radius bins (dotted line). This factor alone yields an odds ratio of 4100 for Kepler-62c. The occurrence rate of planets the size of Kepler-62c would have to be more than 11 times smaller than our extrapolation to yield a confidence level less than 99.7% (green line).

The evaluation of the Kepler-62f signal proceeds in a similar manner but with an extrapolation out to longer orbital periods. Figure S14 shows the number of transiting planets per star with a size within the $3\sigma$ error bars of Kepler-62f, in different period bins (solid black line), the extrapolation to longer orbital periods (black dotted line), and the occurrence rates that would yield a validation at the 99.7% confidence level (green). Also plotted, for comparison, is the number of transiting/eclipsing blends, or false positives (red). Note that the distribution of possible blends is computed using a similar period extrapolation for Neptune-sized companions since they are a significant contributor to the pool of possible blend scenarios. The analysis yields an odds ratio of 5900 for Kepler-62f. Even if the occurrence rate of planets with 200-300 day orbital periods is 15.9 times smaller than the occurrence rate of planets with 100-day orbital periods, Kepler-62f would still be validated with 99.7% confidence.

There is no indication from radial velocity surveys or from the trends in the occurrence rates that are emerging from the Kepler data (*56*) that there is a cliff in the occurrence rates between Earth-size and Mars-size planets or between 100-day and 300-day orbital periods. In the discussions that ensue, we assume that the planet interpretation is by far the most likely interpretation to explain the transits of Kepler-62c & -62f even under the most conservative of assumptions; i.e. that these are validated planets with > 99.7% confidence.

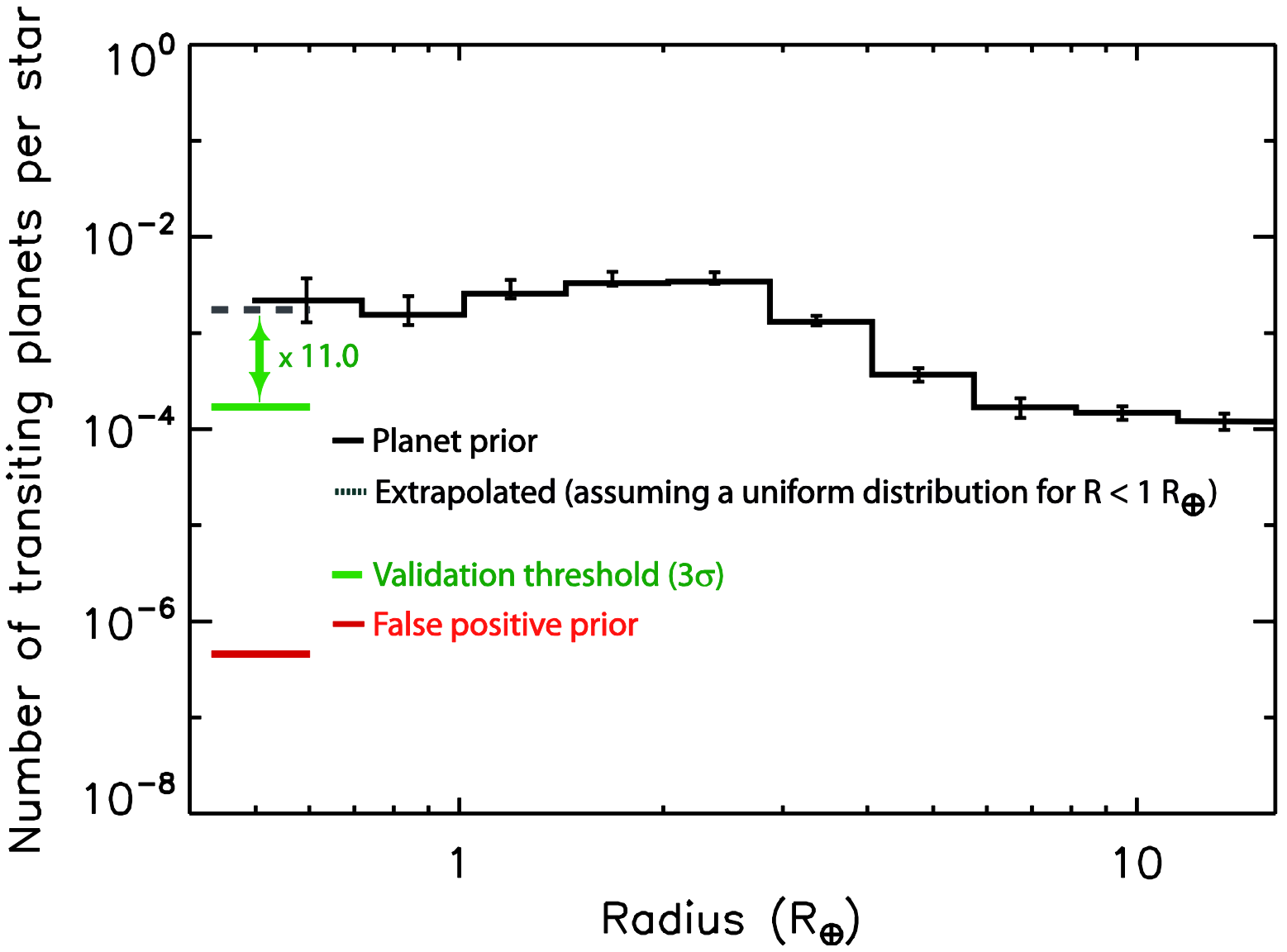

**Fig. S13**. BLENDER results for Kepler-62c. The expected number of planets of similar size and period is 11 times higher than necessary for a 3σ validation.

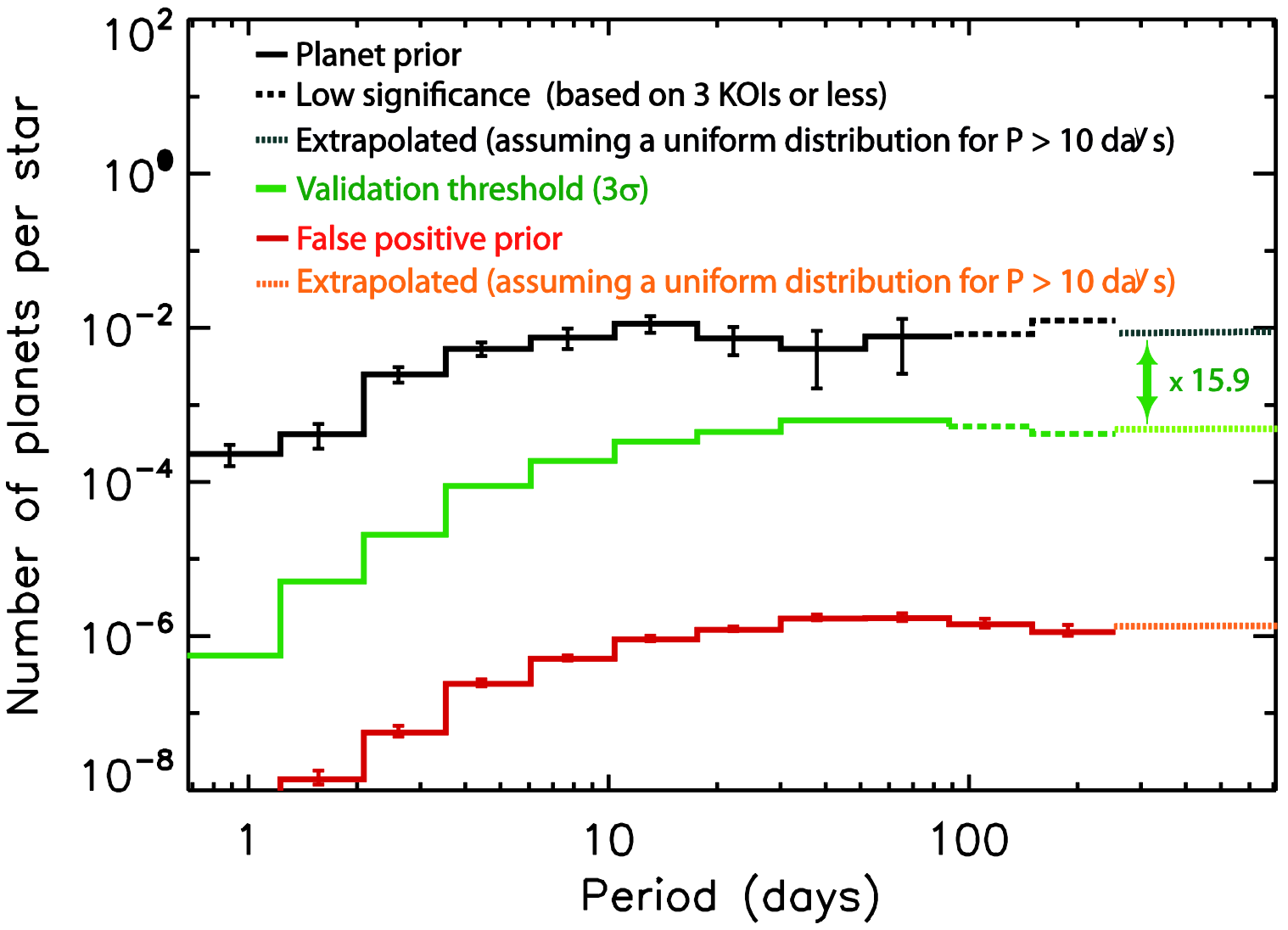

**Fig. S14.** BLENDER results for Kepler-62f. The expected number of planets of similar size and period is 15.9 times higher than necessary for a 3σ validation.

There is also a 0.2% chance that the planets orbit a widely space binary composed of two K2V stars and therefore the planets are √2 larger in radius than shown in Table 1. A twin star could be there (and unseen) if:

a) It is not seen in the most precise AO image (using the limit from the KECK AO of ~0.07" for stars within 2 mag).

b) It does not show a second set of lines in the spectra, nor does it induce a detectable global drift in the RV observations within the span of the observations.

c) It does not make the system dynamically unstable. The outer planet (-62f) sets a minimum distance for the companion star.

d) It does not show a discrepancy with the observed colors of the target star.
To a conservative estimate, we assumed that both the planets and the binary are in circular orbits. The multiplicity of K-stars is estimated to be 40% for K stars (*57*), and their distributions of period and eccentricity were used. Random positions were assigned to the companion stars in a large Monte-Carlo simulation. Dynamic stability was assessed from (*55*).

**8. Summary**

An analysis of the spectrum of Kepler-62 (KIC 9002278) shows it to be a slowly rotating, middle-age K2 dwarf. Searches for confounding stars in the Kepler aperture using active optics, warm-Spitzer, and high-SNR Keck spectra detect no stars bright enough to mimic the transit patterns, but cannot rule out very faint stars. Analysis of the Kepler measurements of image motions by each planet is consistent with these observations. Modeling results from the Blender program estimate the odds that any one of the five planets is a false-positive event to be less than 1 in 5000. There is also a 0.2% chance that the planets orbit a widely-spaced binary system composed of two nearly-identical K2V stars and therefore the radii of the planets are $\sqrt{2}$ larger in radius than shown in Table 1. MCMC modeling of the stellar and planetary parameters provides stable solutions for Kepler-62b, -62c, -62d, -62e, and -62f with sizes of 1.3, 0.54, 1.97, 1.61, and 1.41 $R_{\oplus}$ with orbital periods of 5.7, 12.4, 18.2, 122.4, and 267.3 days, respectively. Kepler-62f is currently the smallest planet detected in the HZ by the Kepler Mission.